\documentclass[12pt]{article}
\usepackage{amsmath}
\usepackage{mathtools}
\usepackage{amssymb} 
\usepackage{amsthm}
\usepackage{bbm}
\usepackage{graphicx,epsf}
\usepackage{epstopdf}
\usepackage{enumerate}
\usepackage{caption}
\usepackage{natbib}
\usepackage{url} 
\usepackage{bm}
\usepackage{color}

\newcommand{\blind}{0}

\newtheorem{definition}{Definition}

\newtheorem{proposition}{Proposition}

\newcommand{\med}{\mbox{median}}
\newcommand{\hmed}{\mbox{hmed}}
\newcommand{\mad}{\mbox{mad}}
\newcommand{\hmad}{\mbox{hmad}}
\newcommand{\IF}{\mbox{IF}}
\DeclareMathOperator*{\argmin}{arg\,min}

\newcommand{\eps}{\varepsilon}
\newcommand{\gs}{\geqslant}
\newcommand{\ls}{\leqslant}
\newcommand{\hmu}{\hat{\mu}}
\newcommand{\hs}{\hat{\sigma}}

\newcommand{\bzero}{\boldsymbol 0}

\newcommand{\bX}{\bm X}

\begin{document}

\def\spacingset#1{\renewcommand{\baselinestretch}%
{#1}\small\normalsize} \spacingset{1}


\if0\blind
{
  \title{\bf A generalized spatial sign\\ 
	           covariance matrix}		
  \author{Jakob Raymaekers and Peter J. 
	  Rousseeuw 
		\vspace{.25cm} \\
		Department of Mathematics, KU Leuven, Belgium}
  \maketitle
} \fi

\if1\blind
{
  \bigskip
  \bigskip
  \bigskip
  \begin{center}
	  {\LARGE\bf A generalized spatial sign covariance matrix}			
\end{center}
  \medskip
} \fi

\bigskip
\begin{abstract}
The well-known spatial sign covariance matrix (SSCM)
carries out a radial transform which moves all data 
points to a sphere, followed by computing the classical 
covariance matrix of the transformed data.
Its popularity stems from its robustness to outliers,
fast computation, and applications to correlation
and principal component analysis.
In this paper we study more general radial functions.
It is shown that the eigenvectors of the generalized 
SSCM are still consistent and the ranks of the
eigenvalues are preserved.
The influence function of the resulting scatter
matrix is derived, and it is shown that its breakdown
value is as high as that of the original SSCM.
A simulation study indicates that the best results are 
obtained when the inner half of the data points are not 
transformed and points lying far away are moved to the 
center.
\end{abstract}

\vskip0.3cm
\noindent%
robust location and scatter. \\

\spacingset{1.45} 
\section{Introduction}
\label{sec:intro}

Robust estimation of the covariance (scatter) matrix 
is an important and challenging problem.
Over the last decades, many robust estimators for the 
covariance matrix have been developed.
Many of them possess the attractive property of affine 
equivariance, meaning that when the data are subjected 
to an affine transformation the estimator will 
transform

\newpage
\noindent accordingly.
However, all highly robust affine equivariant scatter
estimators have a combinatorial time complexity.
Other estimators posses the less restrictive property 
of orthogonal equivariance. 
This means that the estimators commute with orthogonal 
transformations, which are characterized by orthogonal 
matrices and include rotations and reflections. 

The most well-known orthogonally equivariant scatter 
estimator is the spatial sign covariance matrix (SSCM)
proposed independently by \cite{Marden:SSprcomp} and 
\cite{Visuri:Rank} and studied in more detail by 
\cite{Magyar:SSCM} and \cite{Durre:SSCM,Durre:SSeigval} 
among others. 
The estimator computes the regular covariance matrix 
on the {\it spatial signs} of the data, which are the 
projections of the location-centered datapoints on 
the unit sphere. 
Somewhat surprisingly, this transformation yields a 
consistent estimator of the eigenvectors of the true 
covariance matrix \citep{Marden:SSprcomp} under 
relatively general conditions on the underlying 
distribution. 
Of course the eigenvalues are different from the 
eigenvalues of the true covariance matrix, but 
\cite{Visuri:Rank} have shown that the order of the 
eigenvalues is preserved. 
We build on this idea by illustrating that the SSCM is 
part of a larger class of orthogonally equivariant 
estimators, all of which estimate the eigenvectors of 
the true covariance matrix and preserve the order of 
the eigenvalues.

The SSCM is easy to compute, and has been used 
extensively in several applications. 
The most common use of the SSCM is probably in the 
context of (functional) {\it spherical PCA} as 
developed by 
\cite{Locantore:prcomprob}, \cite{Visuri:SSapp}, 
\cite{Croux:SignandRank} and \cite{Taskinen:prcomprob}. 
Like classical PCA, spherical PCA aims
to find a lower dimensional subspace that captures most
of the variability in the data. 
After centering the data, spherical PCA projects 
the data onto the unit (hyper)sphere before searching 
for the directions of highest variability. 
This projection gives all data points the same weight 
in the estimation of the subspace, thereby limiting 
the influence of potential outliers. 
The directions (`loadings') of spherical PCA thus
correspond to the eigenvectors of the SSCM scatter
matrix.
The corresponding scores are usually taken to be the 
inner products of the loading vectors with the 
original (centered) data points, not with the
projections of the data points on the sphere.
Some concrete applications of spherical PCA are
about the shape of the cornea in ophthalmology
as analyzed by \cite{Locantore:prcomprob}, and 
for multichannel signal processing as illustrated 
in \cite{Visuri:Rank}.

In addition to spherical PCA,
there also has been a lot of recent research on the 
use of the SSCM for constructing robust correlation 
estimators
\citep{Durre:SScorr,Durre:SScor2step,Durre:SScorrapp}.
The main focus of this work is on results including 
asymptotic properties, the eigenvalues, and the 
influence function which measures robustness. 
A third application of the SSCM is its use as an 
initial estimate for more involved robust scatter 
estimators \citep{Croux:kSSCM,Hubert:DetMCD}. 
The SSCM is particularly well-suited for this task 
as it is very fast and highly robust against 
outlying observations and therefore often yields a 
reliable starting value.
Another application of the SSCM is to testing
for sphericity \citep{Sirkia:signtest}, which
uses the asymptotic properties of the SSCM in order
to assess whether the underlying distribution of 
the data deviates substantially from a spherical 
distribution.
\cite{Serneels:SSprep} use the spatial sign 
transform as an initial preprocessing step in order
to obtain a robust version of partial least
squares regression. 
Finally, \cite{Boente:operator} study
SCCM as an operator for functional data analysis.

The next section introduces a generalization of the
SSCM and studies its properties. 
Section \ref{sec:sim} compares the performance of 
several members of this class in a small simulation 
study, and Section \ref{sec:concl} concludes.
All proofs can be found in the Appendix.
 
\section{Methodology}
\label{sec:meth}

\subsection{Definition}
\begin{definition}
Let $X$ be a $p$-variate random variable 
and $\mu$ a vector serving as its center. 
Define the \textbf{generalized spatial sign covariance 
matrix} (GSSCM) of $X$ by
\begin{equation}\label{eq:GSSCM}
S_{g_{X}}(X) = E_{F_X}[g_X(X- \mu )g_X(X -\mu )^T]\;,
\end{equation}
where the function $g_X$ is of the form 
\begin{equation}\label{eq:gfunc}
\displaystyle
	g_X(t) = t \; \xi_X(||t||)\;,							
\end{equation}
where we call $\xi_X: \mathbb{R}^+ \to \mathbb{R}^+$ 
the radial function and $||\cdot||$ is the Euclidean 
norm.
\end{definition}
Note that the form of $g_X$ in (\ref{eq:gfunc}) 
precisely characterizes an orthogonally equivariant 
data transformation as shown by
\cite{Hampel:IFapproach}, p. 276.
Also note that the regular covariance matrix
corresponds to $\xi_{X}(r) = 1$, and that 
$\xi_{X}(r) = 1/r$ yields the SSCM.

For a finite data set $\bX = \{x_1, \ldots, x_n\}$ 
the GSSCM is given by
\begin{equation}\label{eq:Sg}
S_{g_{\bX}}(\bX) = \frac{1}{n}\sum_{i=1}^{n}
  {\xi^2_{\bX}(||x_i- T(\bX)||)
	(x_i- T(\bX))(x_i- T(\bX))^T}
\end{equation}
where $T$ is a location estimator.
Note that the SSCM gives the $x_i$ with 
$||x_i - T(\bX)||<1$ 
a weight higher than 1, but in general this is not 
required. In fact, the other functions we will propose 
satisfy $\xi_{\bX}(r) \ls 1$ for all $r$.

In the above definitions, we added the subscript $X$ or 
$\bX$ to the functions $g$ and $\xi$ to indicate 
that they can depend on $X$ or $\bX$. 
In what follows we will drop these subscripts to ease 
the notational burden. 
We will study the following functions $\xi$:
\begin{enumerate}
\item Winsorizing (Winsor): \begin{equation}
\displaystyle
  \xi(r) = 
	\begin{cases}
	    1 & \mbox{ if } r \ls Q_2 \\
	Q_2/r & \mbox{ if } Q_2 < r \;.\\
   \end{cases}
	 \label{eq:Winsor}
\end{equation}
\item Quadratic Winsor (Quad): \begin{equation}
\displaystyle
  \xi(r) = 
	\begin{cases}
	  1 & \mbox{ if } r \ls Q_2 \\
		Q_2^2/r^2 & \mbox{ if } Q_2 < r \;.\\
   \end{cases}
	 \label{eq:Quad}
\end{equation}
\item Ball:
\begin{equation}
\displaystyle
  \xi(r) = 
	\begin{cases}
	 1 & \mbox{ if } r \ls Q_2 \\
	 0 & \mbox{ if } Q_2 < r \;.
   \end{cases}
	 \label{eq:Ball}		
\end{equation}
\item Shell: 
\begin{equation}
\displaystyle
  \xi(r) = 
	\begin{cases}
	 0 & \mbox{ if } r < Q_1 \\
	 1 & \mbox{ if } Q_1 \ls r \ls Q_3 \\
   0 & \mbox{ if } Q_3 < r \;.
   \end{cases}
	 \label{eq:Shell}		
\end{equation}
\item Linearly Redescending (LR):
\begin{equation}
\displaystyle
  \xi(r) = 
	\begin{cases}
	 1   & \mbox{ if } r \ls Q_2 \\
	 (Q_3^* - r)/(Q_3^* - Q_2) &
		\mbox{ if } Q_2 < r \ls Q_3^* \\
   0   & \mbox{ if }  Q_3^* < r \;.
   \end{cases}
	 \label{eq:LR}
	\end{equation}
\end{enumerate}
The cutoffs $Q_1$, $Q_2$, $Q_3$ and $Q_3^*$ depend 
on the Euclidean distances 
$||x_i-T(\bX)||$ by
\begin{align*}
Q_1 &= \left(\hmed_i(||x_i-T(\bX)||^{\frac{2}{3}}) - 
    \hmad_i(||x_i-T(\bX)||^{\frac{2}{3}})
		\right)^{\frac{3}{2}}	\\
Q_2 &= \left(\hmed_i(||x_i-T(\bX)||^{\frac{2}{3}})
    \right)^{\frac{3}{2}}
		= \hmed_i(||x_i-T(\bX)||)	\\
Q_3 &= \left(\hmed_i(||x_i-T(\bX)||^{\frac{2}{3}}) +
    \hmad_i(||x_i-T(\bX)||^{\frac{2}{3}})
		\right)^{\frac{3}{2}}\\
Q_3^* &= \left(\hmed_i(||x_i-T(\bX)||^{\frac{2}{3}}) + 
    1.4826\;\hmad_i(||x_i-T(\bX)||^{\frac{2}{3}})
		\right)^{\frac{3}{2}},
\end{align*}
where $\hmed$ and $\hmad$ are variations on the
median and median absolute deviation given by 
the order statistic
$\hmed(y_1,\ldots,y_n) = y_{(h)}$ and
$\hmad(y_1,\ldots,y_n) = \hmed_i|y_i-hmed_j(y_j)|$ 
where
$h = \left\lfloor\frac{n+p+1}{2} \right\rfloor$\;. 
The $\frac{2}{3}$ power in these formulas is the
Wilson-Hilferty transformation \citep{WilsHilf} to
near normality. In Section \ref{A:dd2} 
of the Appendix
it is verified that this transformation brings the 
above cutoffs close to the theoretical ones, which are 
quantiles of a convolution of Gamma random variables
with different scale parameters.  

\begin{figure}[!ht]
\centering
\includegraphics[width=0.5\textwidth]
                {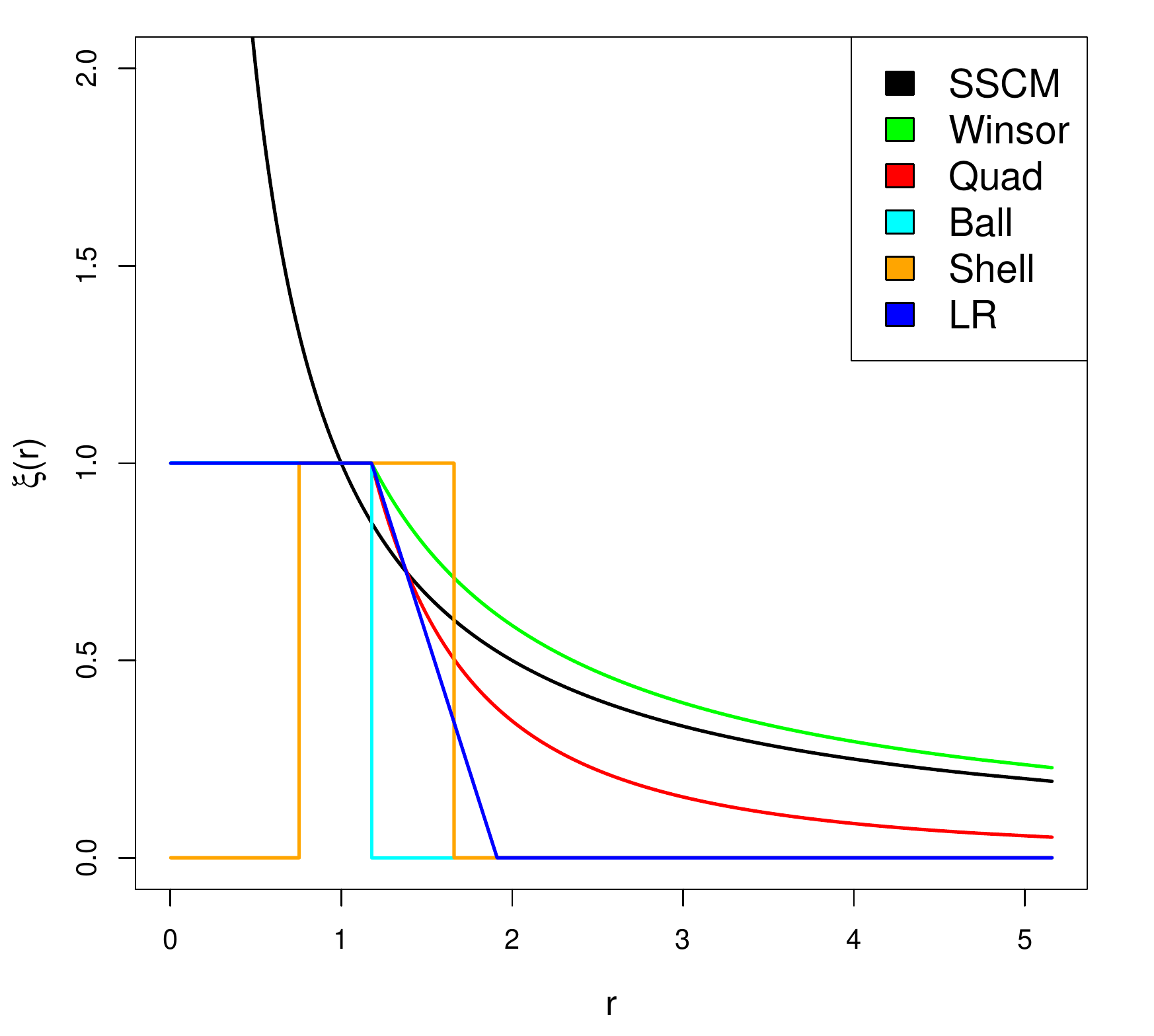}
		\vskip-0.2cm					
\caption{Radial functions $\xi$}
		\label{fig:weightf}
\end{figure}

Figure \ref{fig:weightf} shows the above functions $\xi$ 
and that of the SSCM for distances whose square follows
the $\chi_2^2$ distribution.
The $\xi$ of the SSCM is the only one which upweights
observations close to the center. 
The Winsor $\xi$ and its square have a similar shape, 
but the latter goes down faster.
The Ball and Shell $\xi$ functions are both designed to 
give a weight of 1 to half (in fact, $h$) of the data 
points and 0 to the remainder, to make them comparable. 
Ball does this by giving a weight of 1 to the $h$ points 
with the smallest distances. 
Shell is inspired by the idea of Rocke to both downweight 
observations with very high and very low distances from 
the center \citep{Rocke:Sestimator}.
The Linearly Redescending $\xi$ is a compromise between 
the Ball and the Quad $\xi$ functions.

\subsection{Preservation of the eigenstructure}
In what follows, we assume that the distribution $F_X$ 
of $X$ has an elliptical density with center zero and
that its covariance matrix $\Sigma = E_{F_X}[X X^T]$
exists.
Therefore, $X$ can be written as $X=UDZ$ where $U$ is 
a $p \times p$ orthogonal matrix, $D$ is a $p \times p$ 
diagonal matrix with strictly positive diagonal 
elements, and $Z$ is a $p$-variate random variable which 
is spherically symmetric, i.e. its density is of the 
form $f_Z(z) \sim w(||z||)$ where $w$ is a decreasing 
function. Assume w.l.o.g. that the covariance matrix
of $Z$ is $I_p$.
The following proposition says that $S_g(X)$ has the same 
eigenvectors as $\Sigma$ and preserves 
the ranks of the eigenvalues.

\begin{proposition}\label{prop:eigvec}
Let $X=UDZ$ be a $p$-variate random variable as described 
above, with $D = \mbox{diag}(\delta_1,\ldots,\delta_p)$ 
where $\delta_1 \gs \ldots \gs \delta_p >0$. 
Assume that the covariance matrix 
$S_g = E_{F_X}[g(X)g(X)^T]$ of $g(X)$ exists.
Then $\Sigma$ and $S_g$ can be diagonalized as
\[ 
\Sigma = U \Lambda U^T \;\;\; \mbox{ and }
   \;\;\; S_g = U \Lambda_g U^T
\]
where $\Lambda = \mbox{diag}(\lambda_1,\ldots,\lambda_p)$ 
with $\lambda_j = \delta_j^2$ and
$\Lambda_g = \mbox{diag}(\lambda_{g,1},\ldots,\lambda_{g,p})$ 
with 
$\lambda_{g,1} \gs \ldots \gs \lambda_{g,p} >0$
and $\lambda_j = \lambda_{j+1} \Leftrightarrow 
 \lambda_{g,j} = \lambda_{g,j+1}$\;.
\end{proposition}

This proposition justifies the generalized SSCM approach. 

\subsection{Location estimator}

So far we have not specified any location estimator $T$.
For the SSCM the most often used location 
estimator is the {\it spatial median}, see e.g. 
\cite{Gower:Smedian} and \cite{Brown:Smedian}, which we 
denote by $T_0$. 
The spatial median of a dataset $\bX=\{x_1,\ldots,x_n\}$
is defined as
\begin{equation*}
T_0(\bX) = \argmin_{\theta}{
      \sum_{i=1}^{n}{||x_i-\theta||}}\;.
\end{equation*}

In order to improve its robustness against a substantial
fraction of outliers we propose to use the 
\textit{k-step least trimmed squares (LTS) estimator}.
The LTS method was originally proposed in regression
\citep{Rousseeuw:LMS}, and for multivariate location 
it becomes
\begin{equation*}
T_{\mbox{LTS}}(\bX) = \argmin_{\theta} {\sum_{i=1}^{h} 
  ||x_\bullet-\theta||^2_{(i)}}
\end{equation*}
where the subscript $(i)$ stands for the i-th
smallest squared distance.
(Without the square this becomes the least trimmed 
absolute distance estimator studied in 
\cite{Chatzinakos:LTED}.)
For the multivariate location LTS the C-step of
\citep{Rousseeuw:FastMCD} simplifies to

\begin{definition} (C-step) 
Fix $h = \left\lfloor (n+1)/2 \right\rfloor$. Given 
a location estimate $T_{j-1}(\bX)$ we take the set
$I_j = \{i_1,\ldots,i_h\} \subset \{1,\ldots,n\}$ 
such that $\{||x_i- T_{j-1}(\bX)||;\; i \in I_j\}$ 
are the $h$ smallest distances in the set 
$\{||x_{i}- T_{j-1}(\bX)||;\; i=1,\ldots,n\}$.
The C-step then yields
\begin{equation*}
 T_j(\bX) = \frac{1}{h}\sum_{i \in I_{j-1}}x_i\;.
\end{equation*}
\end{definition}
The C-step is fast to compute, and guaranteed to 
lower the LTS objective.
The k-step LTS is
then the result of  $k$ successive C-steps starting
from the spatial median $T_0(\bX)$\;.

It is also possible to avoid the estimation of location 
altogether, by calculating the GSSCM on the $O(n^2)$ 
pairwise differences of the data points. This approach is 
called the ``symmetrization'' of an estimator,
but is more computationally intensive. 
\cite{Visuri:Rank} studied the symmetrized SSCM and called 
it Kendall's $\tau$ covariance matrix.

\subsection{Robustness properties}

A major reason for the SSCM's popularity is its robustness 
against outliers.
Robustness can be quantified by the influence function and 
the breakdown value. We will study both for the GSSCM.

The influence function (\cite{Hampel:IFapproach}) 
quantifies the effect of a small amount of contamination 
on a statistical functional $T$. 
Consider the contaminated distribution 
$F_{\eps,z}=(1-\eps)F+\eps \Delta(z)$, 
where $\Delta(z)$ is the distribution that puts all its 
mass in $z$. 
The influence function of $T$ at $F$ is then given by
\begin{equation*}
\mbox{IF}(z,T,F)=\lim_{\eps \to 0}
    \frac{T(F_{\eps,z})-T(F)}{\eps}
		= \frac{\partial}{\partial \eps}
	  \;T(F_{\eps,z})\bigg|_{\eps=0}.
\end{equation*}

For the generalized SSCM class we obtain
the following result:
\begin{proposition}\label{prop:IF}
Denote $S_g(F) = \Xi_g$ and let $\mu=0$ in 
\eqref{eq:GSSCM}.
The influence function of $S_g$ at the distribution 
$F$ is given by:
\begin{equation}\label{eq:IF}
\IF(z, S_g, F) = \left. \frac{\partial}{\partial \eps} 
   S(F_{\eps, z}) \right|_{\eps = 0}  =
	 g(z)g(z)^T - \Xi_g + \left. 
	\frac{\partial}{\partial \eps} 
	\int{g_{\eps}(X)g_{\eps}(X)^T dF(X)} 
	\right|_{\eps = 0} \; .
\end{equation}
\end{proposition}
If $g$ does not depend on $F$, the last term of 
(\ref{eq:IF}) vanishes. 
For example, for $g(t) = t$, we retrieve the IF of the 
classical covariance matrix 
$\IF(z, \Sigma, F) = zz^T - \Sigma$\;, and for 
$g(t) = t/||t||$ we obtain 
$\IF(z, \mbox{SSCM}, F) = \left(\frac{z}{||z||}\right)
\left(\frac{z}{||z||}\right)^T - \mbox{SSCM}(F)$ in line
with the findings of \cite{Croux:SignandRank}. 
For the GSSCM estimators defined by the functions
\eqref{eq:Winsor}--\eqref{eq:LR} the last term 
of (\ref{eq:IF}) remains, and the expressions 
of their IF can be found in Section \ref{A:IF} 
of the Appendix.

\begin{figure}[!ht]
\centering
\includegraphics[width=0.9\textwidth]{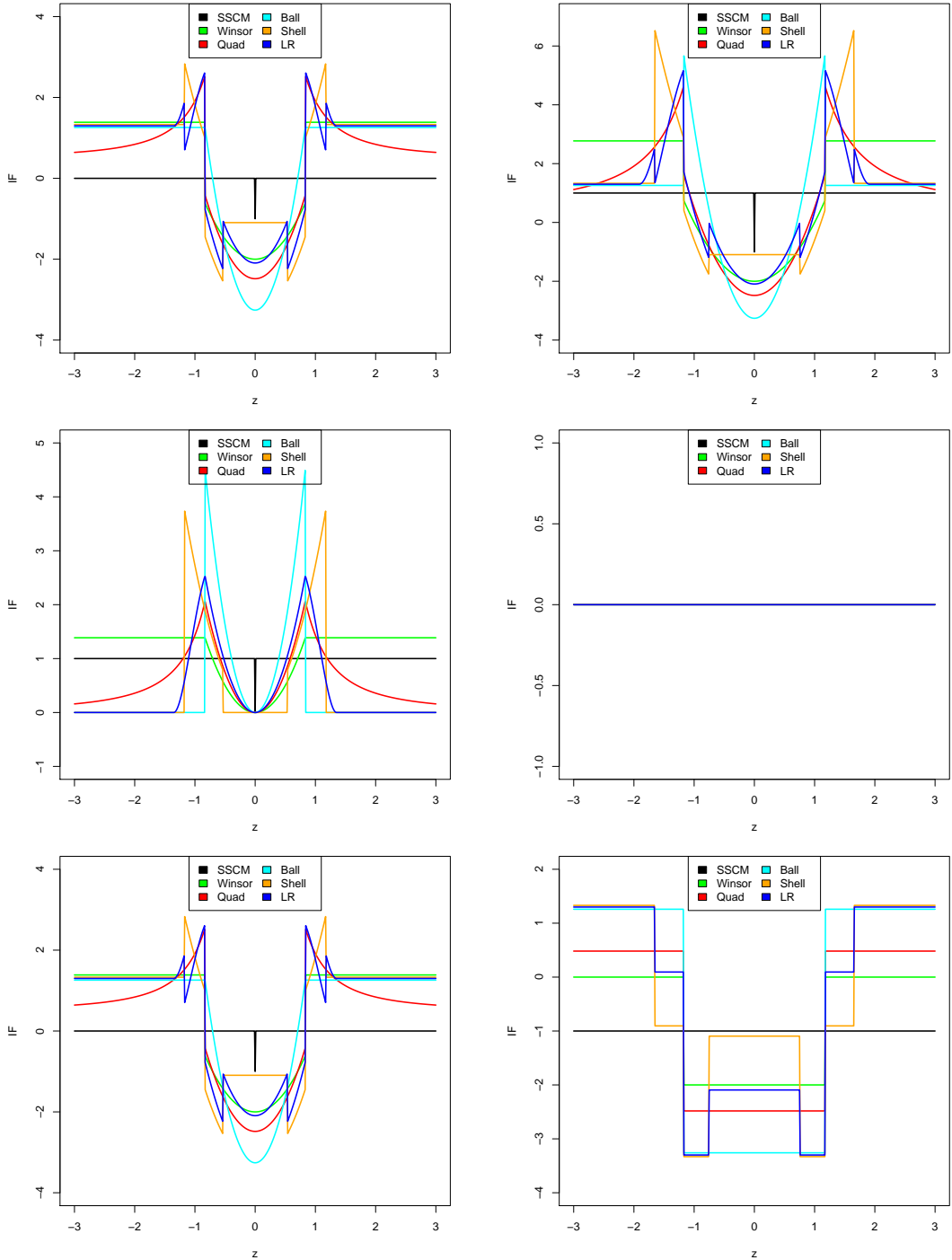} 
\caption{Influence functions of the GSSCM at the bivariate 
standard normal distribution for contamination at $(z,z)$ 
(left) and $(z,0)$ (right). The rows correspond to the
first diagonal element $S_{11}$ (top), the off-diagonal
element $S_{12}$ (middle), and $S_{22}$ (bottom).}
\label{fig:IF}
\end{figure}

In order to visualize the influence function we consider 
the bivariate standard normal case, i.e. 
$F = N(0,I_2)$.
We put contamination at $(z,z)$ or $(z,0)$ for different 
values of $z$ and plot the IF for the diagonal elements 
and the off-diagonal element. 
Note that we cannot compare the raw IFs directly as 
$S_g(F) = \Xi_g = c_gI$ where $c_g = \int{g_1(X)^2dF(X)}$ 
hence $\Xi_g$ is only equal to $I_2$ up to a factor. 
In order to make the estimators consistent for this 
distribution we can divide them by $c_g$\,, and so we 
plot $\IF(z,S_g,F)/c_g$ in Figure \ref{fig:IF}.

The rows in Figure \ref{fig:IF} correspond to the IF of
the first diagonal element $S_{11}$ (top), the off-diagonal
element $S_{12}$ (middle) and the element $S_{22}$ (bottom).
Let's first consider the left part of the figure, which 
contains the IFs for an outlier in $(z,z)$. 
By symmetry, the IFs of the diagonal elements $S_{11}$
and $S_{22}$ are the same here.
In the regions where the function $\xi$ is 1 the IF is
quadratic, like that of the classical covariance.
The diagonal elements of the IF of the SSCM are zero, 
except at $z=0$ where it takes the value $-1$.
The Quad IF is the only one which redescends as $|z|$ 
increases, whereas the others are also bounded but 
stabilize at a value around $1.3$. 
The shape of the IF of the Ball estimator resembles that 
of the univariate Huber M-estimator of scale. 

For the IF of the off-diagonal element $S_{12}$ the 
picture is very different.
All are redescending except for the SSCM and Winsor.
Here it is Winsor whose IF resembles that of Huber's
M-estimator of scale. 
Note that the IFs of the Ball and Shell 
estimators have large jumps at their cutoff values.
The discontinuities in the IFs are due to the fact that 
the cutoffs depend on the median and the MAD of the 
distances $||X||^{2/3}$, as both the median and the MAD 
have jumps in their IF.
 
The right panel of Figure \ref{fig:IF} shows the 
influence functions for an outlier in $(z,0)$. 
In this case the IFs of the diagonal elements 
$S_{11}$ and $S_{22}$ are no longer the same, as the 
symmetry is broken.
The IFs of $S_{11}$ are again quadratic where $\xi = 1$, 
with jumps at the cutoffs. 
Note that these cutoffs are now located at different 
values of $z$, as $||(z,0)|| \neq ||(z,z)||$.
The IF of the off-diagonal element is constant at 0,
indicating that $S_{12}$ remains zero even when there
is an outlier at $(z,0)$.
Finally, for the second diagonal element $S_{22}$ the 
IF of the SSCM is $-1$. This is because adding $\eps$ of
contamination at $(z,0)$ reduces the mass of the 
remaining part of $F$ by $\eps$ which lowers the 
estimated scatter in the vertical direction.
For the other estimators there is an additional effect 
of $(z,0)$ on the cutoffs, which causes the 
discontinuities.

A second tool for quantifying the robustness of an 
estimator is the finite-sample breakdown value 
\citep{Donoho:BDP}.
For a multivariate location estimator $T$ and a 
dataset $\bX$ of size $n$, the breakdown value is the 
smallest fraction of the data that needs to be 
replaced by contamination to make the resulting 
location estimate lie arbitrarily far away from the 
original location $T(\bX)$. More precisely:
\begin{equation*}
\eps^*(T,\bX) = \min{\left\{\frac{m}{n}: 
     \sup_{\bX_m^*}{||T(\bX) - T(\bX^*)||} = 
		\infty \right\}}
\end{equation*}
where $\bX_m^*$ ranges over all datasets obtained 
by replacing any $m$ points of $\bX$ by arbitrary 
points.

For a multivariate estimator of scale $S$, the 
breakdown value is defined as the smallest fraction 
of contamination needed to make an eigenvalue 
of $S$ either arbitrarily large or arbitrarily close 
to zero.
We denote the eigenvalues of $S(\bX)$ by 
$\lambda_1(S(\bX)) \gs \ldots \gs \lambda_p(S(\bX))$. 
The breakdown value of $S$ is then given by:
\begin{equation*}
\eps^*(S,\bX) = \min{\left\{\frac{m}{n}:
  \sup_{\bX_m^*}{\max{\{\lambda_1(S(\bX_m^*)), 
	\lambda_p(S(\bX_m^*))^{-1}\}} = \infty}
	\right\}}\;.
\end{equation*}

For the results on breakdown we assume the following 
conditions on the function $\xi$:
\begin{enumerate}
\item The function $\xi$ takes values in $[0,1]$.
\item For any dataset $\bX$ it holds that
			$\#\{x_i | \;\xi(||x_i-T(\bX)||) = 1 \} \gs
			\left\lfloor \frac{n + p + 1}{2} \right\rfloor$. 
\item For any vector $t$ it holds that 
      $||g(t)|| = ||t|| \xi(||t||)
      \ls \hmed_i(d_i) + 1.4826\,\hmad_i(d_i)$\;.
\end{enumerate}
Note that all functions $\xi$ proposed in
\eqref{eq:Winsor}--\eqref{eq:LR} satisfy these 
assumptions.
The following proposition gives the breakdown value 
of the GSSCM scatter estimator $S_g$.

\begin{proposition}\label{prop:bdvS}
Let $\bX = \{x_1, \ldots, x_n\}$ be a $p$-dimensional 
dataset in general position, meaning that no $p + 1$ 
points lie on the same hyperplane. Also assume that 
the location estimator $T$ has a breakdown value of 
at least $\left\lfloor (n-p+1)/2\right\rfloor/n$\,. 
Then 
\begin{equation*}
 \epsilon^*\left(S_g, \bX\right) = 
 \frac{\left\lfloor (n-p+1)/2\right\rfloor}{n}\;.
\end{equation*}
\end{proposition}

As we would like the GSSCM scatter estimator to attain 
this breakdown value, we have to use a location 
estimator whose breakdown value is at least 
$\left\lfloor (n-p+1)/2\right\rfloor/n$\,. The following
proposition verifies that the k-step LTS estimator 
satisfies this, and even attains the best possible 
breakdown value for translation equivariant location 
estimators.

\begin{proposition}\label{prop:bdvT}
The k-step LTS estimator $T_k$ satisfies
\begin{equation*}
\eps^*(T_k, \bX) =  
  \frac{\left\lfloor (n+1)/2 \right\rfloor}{n}
\end{equation*}
at any p-variate dataset $\bX = \{x_1, \ldots, x_n\}$.
When the C-steps are iterated until convergence 
$(k \to \infty)$, the breakdown value remains the same.
\end{proposition}

\section{Simulation study}
\label{sec:sim}

We now perform a simulation study comparing the GSSCM 
versions \eqref{eq:Winsor}--\eqref{eq:LR}.
As the estimators are orthogonally equivariant, it 
suffices to generate diagonal covariance matrices.
We generate $m=1000$ samples of size $n=100$ from 
the multivariate Gaussian distribution of dimension 
$p=10$ with center 
$\mu = \bzero$ and covariance matrices 
$\Sigma_1 = I_p$ (`constant eigenvalues'),
$\Sigma_2 = \mbox{diag}(10,9, \ldots, 1)$ 
(`linear eigenvalues'), 
and $\Sigma_3 = \mbox{diag}(10^2, 9^2, \ldots, 1)$
(`quadratic eigenvalues'). 
To assess robustness we also add 20\% and 40\% of 
contamination in the direction of the last 
eigenvector, at the point $(0,\ldots,0,\gamma)$  
for several values of $\gamma$.
For the location estimator $T$ in \eqref{eq:Sg}
we used the k-step LTS with $k=5$.

For measuring how much the estimated $\widehat{\Sigma}$ 
deviates from the true $\Sigma$ we use the 
Kullback-Leibler divergence (KLdiv) given by
\begin{equation*}
  \mbox{KLdiv}(\widehat{\Sigma},\Sigma) = 
  \mbox{trace}(\widehat{\Sigma} \Sigma^{-1}) - 
	\log(\det(\widehat{\Sigma} \Sigma^{-1})) - p\;.
\end{equation*}
We also consider the shape matrices
$\widehat{\Gamma} = 
 (\mbox{det}\widehat{\Sigma})^{-1/p}
 \widehat{\Sigma}$
and
$\Gamma = (\mbox{det}\Sigma)^{-1/p}\Sigma$
which have determinant 1, and compute
$\mbox{KLdivshape}(\widehat{\Sigma},\Sigma) :=
\mbox{KLdiv}(\widehat{\Gamma},\Gamma)$\;.
Both the KLdiv and the KLdivshape are then averaged 
over the $m=1000$ replications.

\begin{figure}[!ht]
\centering
\includegraphics[width=0.99\textwidth]
    {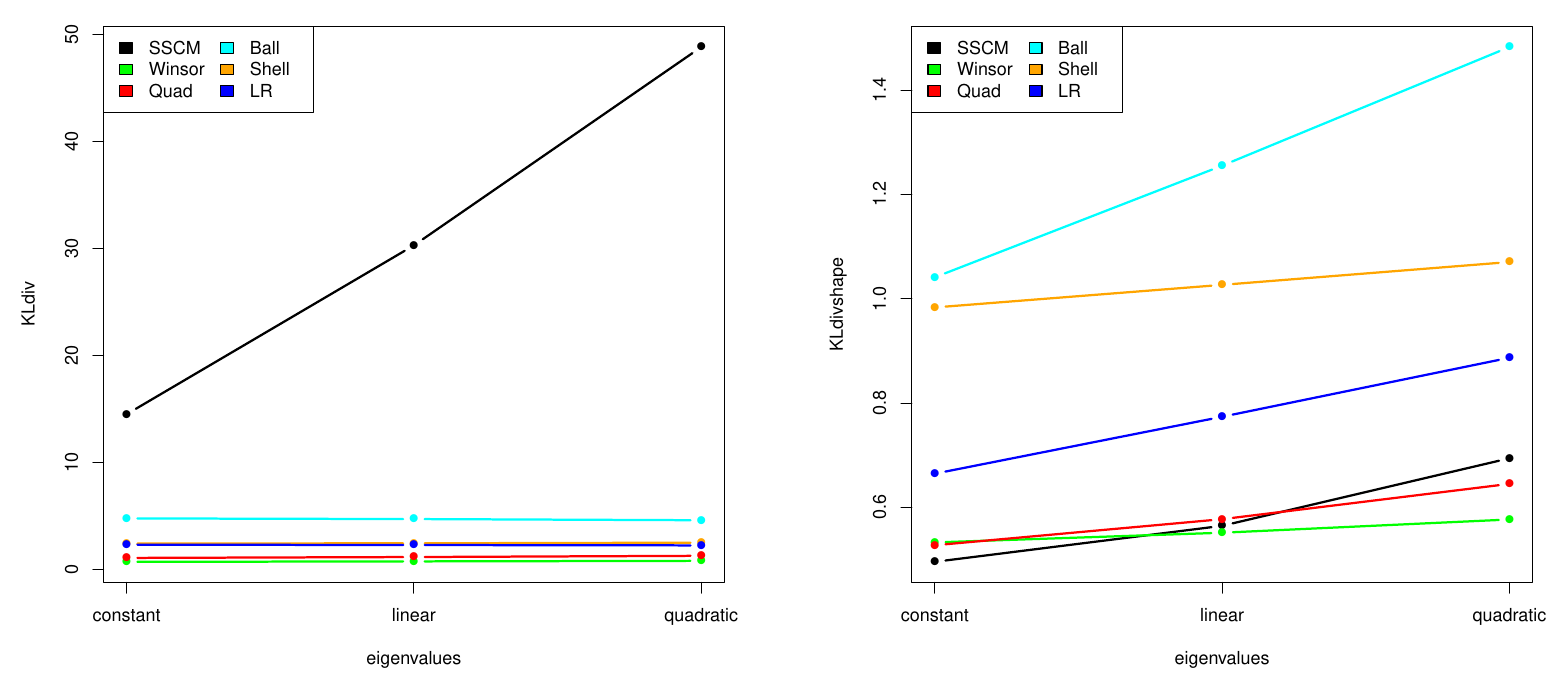}
\caption{Simulation results: KLdiv (left) and KLdivshape 
   (right) for the uncontaminated normal distribution, 
	 with constant, linear and quadratic eigenvalues. }
\label{fig:sim_e0}
\end{figure}

Figure \ref{fig:sim_e0} shows the simulation results 
on the uncontaminated data. 
Looking at KLdiv (left panel) we note that the SSCM
deviates the most from the true covariance matrix
$\Sigma$. 
Among the other choices, Winsor and Quad have the lowest 
bias, followed by LR, Shell, and Ball. 
When looking only at the shape component (right panel),
SSCM performs the best when the distribution is 
spherical (constant eigenvalues), in line with Remark 
3.1 in \citep{Magyar:SSCM}. 
However, it loses this dominant performance once the 
distribution deviates from sphericity. 
Among the other GSSCM methods Winsor performs the best, 
followed by its quadratic counterpart, LR, Shell,
and finally Ball.

\begin{figure}[!ht]
\centering
\includegraphics[width=0.87\textwidth]
	{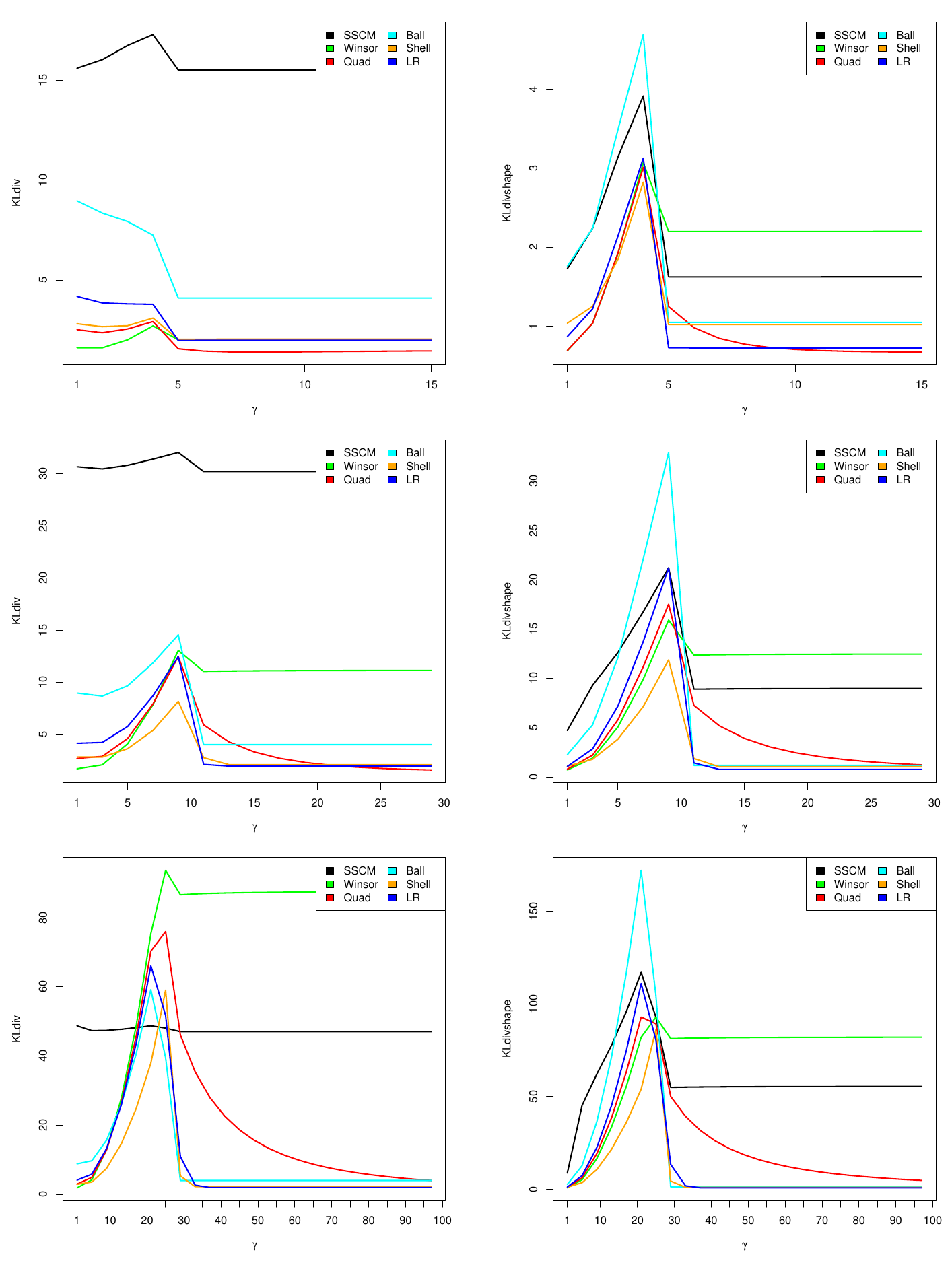} 
\caption{Simulation results: KLdiv (left) and 
  KLdivshape (right) for the normal distribution with 
	constant (top), linear (middle) and quadratic 
	(bottom) eigenvalues and 20\% of contamination.
  The outliers were placed at the point 
	$(0,\ldots,0,\gamma)$.}
\label{fig:sim_e02}
\end{figure}

\begin{figure}[!ht]
\centering
\includegraphics[width=0.87\textwidth]
	{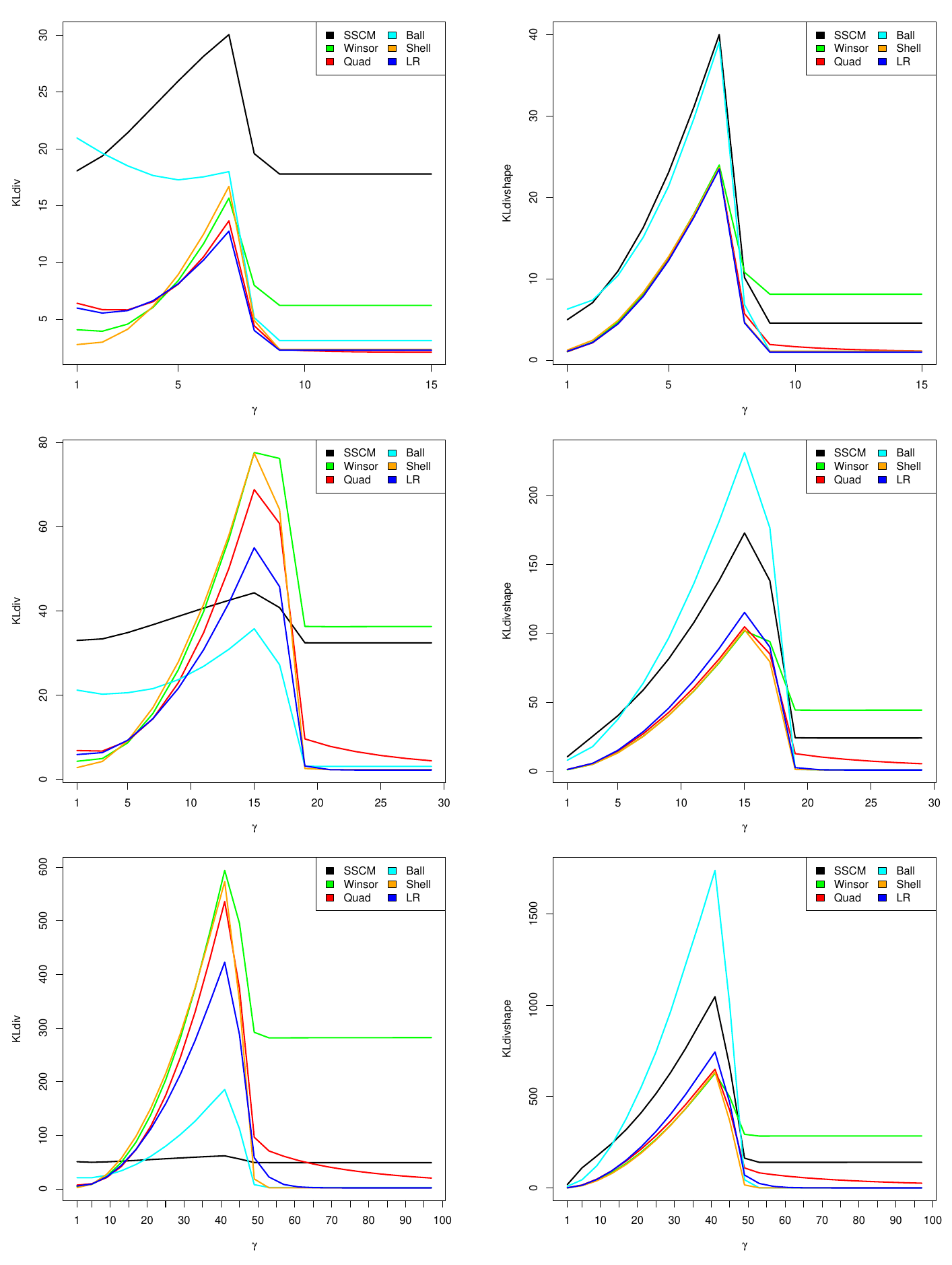} 
\caption{Simulation results: KLdiv (left) and KLdivshape 
  (right) for the normal distribution with constant (top), 
	linear (middle) and quadratic (bottom) eigenvalues and 
	40\% of point contamination. }
\label{fig:sim_e04}
\end{figure}

The result for the simulation with 20\% of point 
contamination is presented in Figure \ref{fig:sim_e02}.
All plots are as a function of $\gamma$, which indicates
the position of the outliers.
In the left panel (KLdiv) the SSCM has a large bias. 
The Winsor GSSCM, which did very well in the uncontaminated 
setting, now has a disappointing performance 
when the eigenstructure becomes more challenging with 
linear or quadratic eigenvalues. Quad performs a lot 
better, but also suffers under quadratic eigenvalues. 
LR and Shell perform the best here, followed by Ball.
Their redescending nature helps them for far outliers.
The conclusions for the shape component (right panel)
are largely similar, except that Winsor and especially
Ball look worse here.

The simulation results for 40\% of contamination are 
shown in Figure \ref{fig:sim_e04}.
The KLdiv plots on the left indicate that the SSCM 
performs poorly for constant and linear eigenvalues,
and looks better for quadratic eigenvalues but not
when $\gamma$ is large (far outliers). 
Winsor performs badly for linear and quadratic
eigenvalues, whereas Quad does much better.
Ball looks okay except for relatively small $\gamma$.
LR and Shell perform the best for both small and
large $\gamma$, and are okay for intermediate
$\gamma$.
When estimating the shape component (right panels)
SSCM and Winsor have the worst performance overall,
whereas Ball also does poorly for small to 
intermediate $\gamma$.
LR and Shell are the best picks here. Quad does
almost as well, but redescends more slowly.

\section{Application: principal component analysis}

We analyze a multivariate dataset from a study by
\cite{Reaven:diabetes}.
The dataset contains 5 numerical variables for 109 
subjects, consisting of 33 overt diabetes patients 
and 76 healthy people.
The variables are body weight, fasting plasma glucose, 
area under the plasma glucose curve, area under the 
plasma insulin curve, and steady state plasma glucose 
response. These data were previously analyzed by 
\cite{Mozharovskyi:DDalpha} in the context of 
clustering using statistical data depth, and is 
available in the R package \textit{ddalpha} 
\citep{Pokotylo:DDalphaR} under the tag 
\textit{chemdiab\_2vs3}. Here we analyze the data by 
principal component analysis.
We first standardize the data, as the variables have
quite different scales. Denote the standardized 
observations by $z_i$ for $i \in \{1, \ldots, 109\}$.

We consider the diabetes patients as outliers and would like the PCA subspace to model the variability within the healthy patients. For classical PCA, the PCA subspace corresponds to the linear span of the $k$ eigenvectors (also called `loadings') of the covariance matrix which correspond with the $k$ largest eigenvalues. In similar fashion we can perform PCA based on the GSSCM, by considering the linear span of its $k$ first eigenvectors. We take $k = 3$ components, thereby explaining more than 95 \% of the variance.  

\begin{figure}[!ht]
\centering
\vskip0.1cm
\includegraphics[width=0.495\textwidth]
   {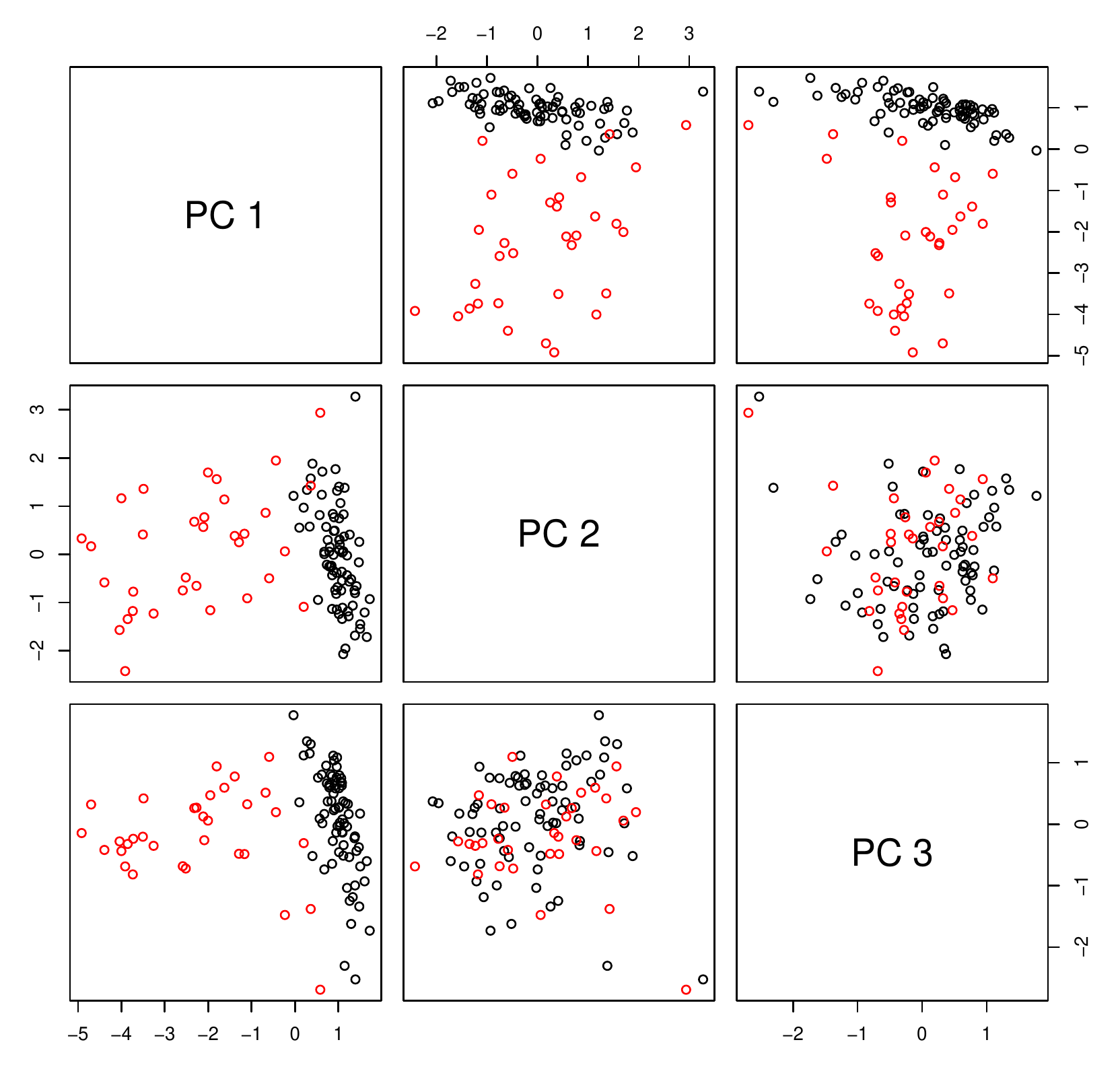}
\includegraphics[width=0.495\textwidth]
   {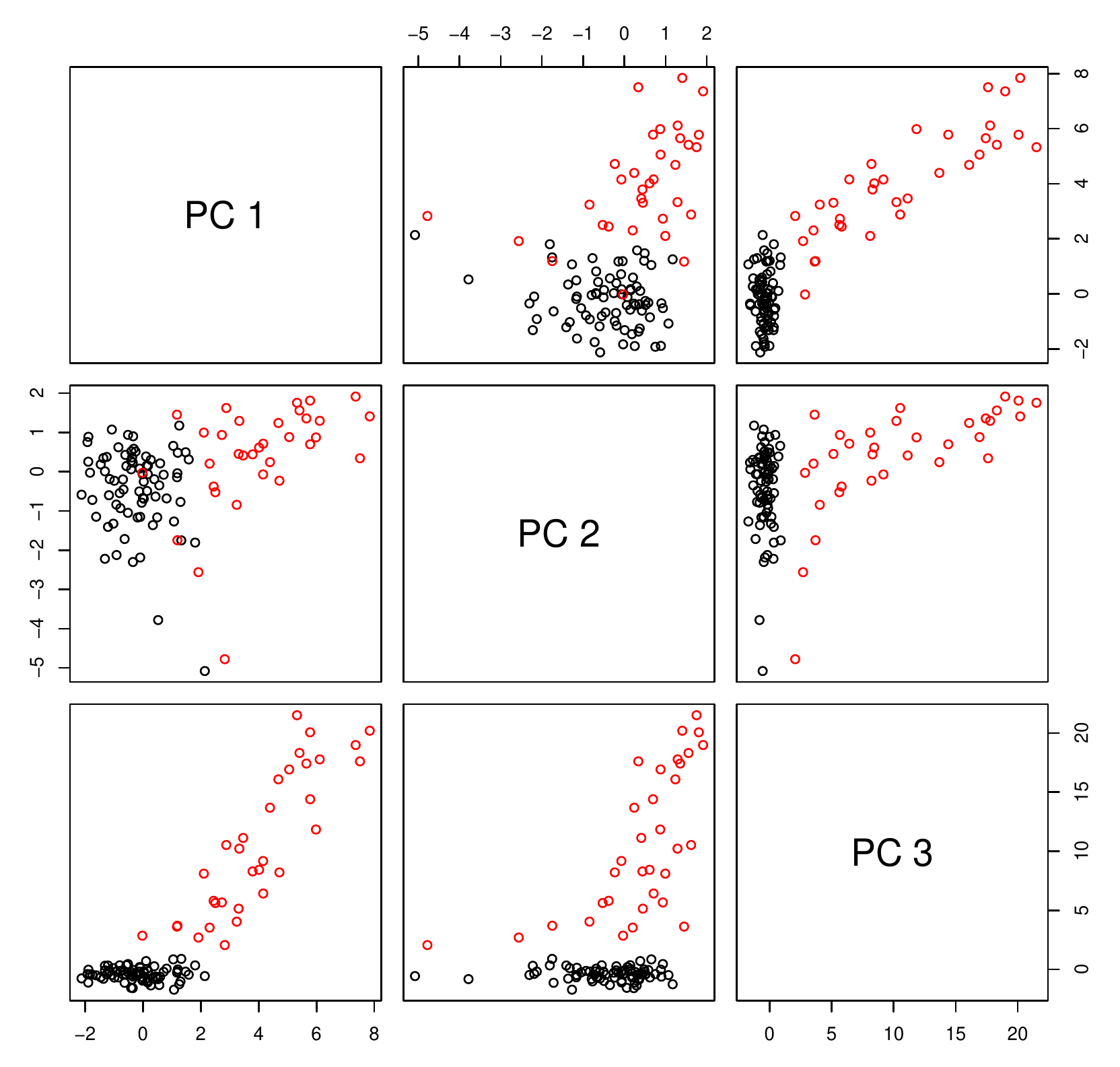} 
\caption{Scores from the 3 first loading vectors of 
         classical PCA (left) and GSSCM PCA (right).}
\label{fig:diabetes_scores}
\end{figure}

Figure \ref{fig:diabetes_scores} shows the scores with respect to the first 3 loadings for classical PCA and GSSCM PCA.
The scores $s_i$ are the projections of the observations $z_i$ onto the PCA subspace, i.e. $s_{i,j} = z_i^T v_j$ where $v_j$ denotes the $j$-th eigenvector. From these plots, it is clear that the first eigenvector of the classical PCA is heavily attracted by the diabetes patients. As a result, the outliers are only distinguishable in their scores with respect to the first principal component. This is very different for the GSSCM PCA, where the principal components seem to fit the healthy patients better, resulting in outlying scores for the diabetes patients with respect to several principal components.
 
\begin{figure}[!ht]
\centering
\vskip0.1cm
\includegraphics[width=0.49\textwidth]
   {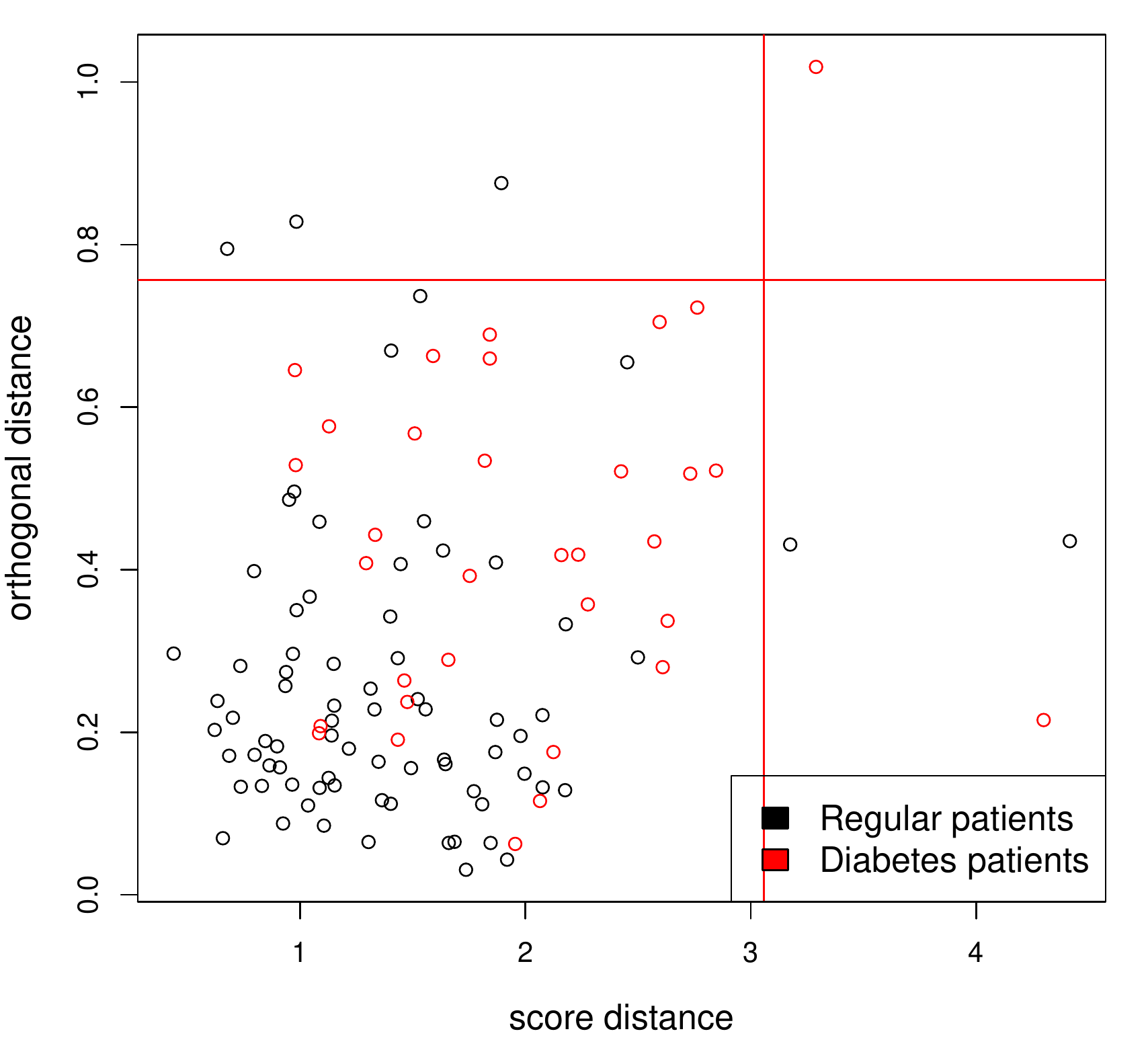} 
\includegraphics[width=0.49\textwidth]
   {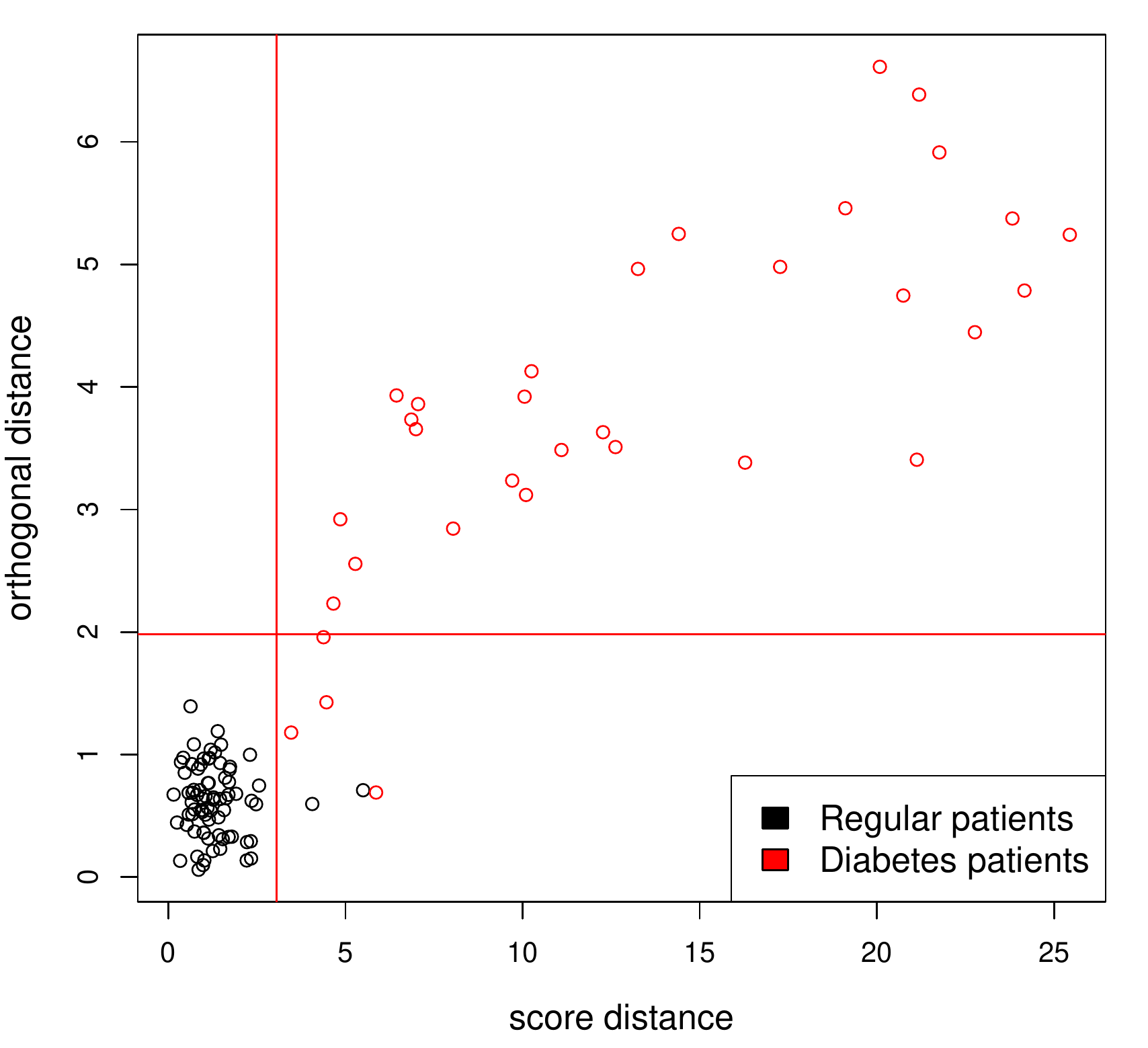} 
\caption{Outlier maps based on classical PCA (left) 
         and GSSCM PCA (right).}
\label{fig:diabetes_DD}
\end{figure}

In addition to the scores plots, the PCA outlier 
map of \cite{Hubert:RobPCA} can serve as a 
diagnostic tool for identifying outliers. 
It plots the orthogonal distance $\mbox{OD}_i$ against
the score distance $\mbox{SD}_i$ for every observation 
$z_i$ in the dataset. 
The score distance of observation $i$ captures the 
distance between the observation and the center of the 
data within the PCA subspace. 
It is given by $\mbox{SD}_i = \sqrt{\sum_{j=1}^{3}
{(s_{ij}/\hat{\sigma}_j})^2}$ where $\hat{\sigma}_j$ 
denotes the scale of the $j$-th scores. For classical 
PCA $\hat{\sigma}_j$ is their standard deviation, 
whereas for GSSCM PCA we take their median absolute 
deviation. 
The orthogonal distance to the PCA subspace is given 
by $\mbox{OD}_i = ||z_i - V s_i||$ where $V$ is 
the $5 \times 3$ matrix containing the 3 eigenvectors 
in its columns. 
Both the score distances and the orthogonal distances
have cutoffs, described in \cite{Hubert:RobPCA}.
Figure \ref{fig:diabetes_DD} shows the outlier maps 
resulting from the classical PCA and the GSSCM PCA. 
Classical PCA clearly fails to distinguish the 
diabetes patients from the healthy subjects. 
In contrast, GSSCM PCA flags most of the diabetes 
patients as having both an abnormally high orthogonal
distance to the PCA subspace as well as having a 
projection in the PCA subspace far away from those
of the healthy subjects.

\section{Conclusions}
\label{sec:concl}
The spatial sign covariance matrix (SSCM) can be seen 
as a member of a larger class called Generalized SSCM 
(GSSCM) estimators in which other radial functions 
are allowed. It turns out that the GSSCM estimators 
are still consistent for the true eigenvectors while
preserving the ranks of the eigenvalues.
Their computation is as fast as the SSCM.
We have studied five GSSCM methods with intuitively 
appealing radial functions, and shown that their 
breakdown values are as high as that of the original 
SSCM. We also derived their influence functions and
carried out a simulation study.

The radial function of the SSCM is $\xi(r) = 1/r$
which implies that points near the center are given
a very high weight in the covariance computation.
Our alternative radial functions give these points
a weight of at most 1, which yields better 
performance at uncontaminated Gaussian data
(Figure \ref{fig:sim_e0}) as well as contaminated 
data
(Figures \ref{fig:sim_e02} and \ref{fig:sim_e04}).
In particular, Winsor is the most similar to SSCM
since its $\xi(r)$ is 1 for the central half of 
the data and $1/r$ for the outer half. 
It performs best for uncontaminated data, but still
suffers when far outliers are present.
It is almost uniformly outperformed by Quad, whose 
$\xi(r)$ is 1 in the central half and $1/r^2$ 
outside it. The influence of outliers on Quad
smoothly redescends to zero.
The other three estimators are hard redescenders
whose $\xi(r)=0$ for large enough $r$.
Among them, the linear redescending (LR) radial
function performed best overall.

A potential topic for further research is to
investigate principal component analysis based
on a GSSCM covariance matrix.\\

\noindent
{\bf Software availability.} R-code for computing
these estimators and an example script
are available from the website
{\it wis.kuleuven.be/stat/robust/software}\,.\\

\noindent
{\bf Acknowledgment.} This research	was supported 
  by projects of Internal Funds KU Leuven.



\clearpage
\pagenumbering{arabic}
%
\appendix
\numberwithin{equation}{section} 
\section{Appendix} \label{sec:A}
\renewcommand{\theequation}
   {\thesection.\arabic{equation}}

Here the proofs of the results are collected.

\subsection{Distribution of Euclidean distances}
\label{A:dd2}

\noindent
\textbf{Exact distribution.}

The exact distribution of the squared Euclidean 
distances $||X||^2$ of a multivariate Gaussian 
distribution with general covariance matrix is
given by the following result:

\begin{proposition}\label{prop:distEuclid}
Let $X \sim N(0, \Sigma)$, and suppose the eigenvalues 
of $\Sigma$ are given by $\lambda_1, \ldots, \lambda_p$\,. 
Then $||X||^2 \sim \sum_{i=1}^{p}{\Gamma\left(\frac{1}{2},
 2\lambda_i\right)}$. For $p\to \infty$ we have
$||X||^2 \xrightarrow{D} N\left(\sum_{i=1}^{\infty}
{\lambda_i}\,, 2\sum_{i=1}^{\infty}{\lambda_i^2}\right)$.
\end{proposition}

\noindent
{\it Proof.}
We can write $X = UDZ$ where $U$ is an orthogonal 
matrix, $D$ is the diagonal matrix with elements 
$\sqrt{\lambda_1},\ldots,\sqrt{\lambda_p}$\;, and
$Z$ follows the $p$-variate standard Gaussian
distribution.
Note that $||X||^2 = ||UDZ||^2 = ||DZ||^2 = 
\sum_{i=1}^{p}{\lambda_i Z_i^2}$ where 
$Z_i^2\sim \chi^2(1)$. 
Therefore, $\lambda_i Z_i^2\sim \Gamma\left(
\frac{1}{2}, 2\lambda_i\right)$ so the distribution 
of $||X||^2$ is a sum of i.i.d. gamma distributions 
with a constant shape of $\frac{1}{2}$ and varying 
scale parameters equal to twice the eigenvalues of 
the covariance matrix.

As $p$ goes to infinity it holds that
\begin{equation*}
||X||^2 \xrightarrow{D} N\left(\sum_{i=1}^{\infty}
{\lambda_i}\,, 2\sum_{i=1}^{\infty}{\lambda_i^2}\right)
\end{equation*}
by the Lyapunov central limit theorem.\\

\noindent
\textbf{Approximate distribution of a sum 
  of gamma variables.}

Proposition \ref{prop:distEuclid} gives the exact
distribution of the squared Euclidean distances
$||X||^2$.
The distribution of a sum of gamma distributions has 
been studied by \cite{Moschopoulos:sumGamma}. 
Quantiles of this distribution can be computed
by the R package \textit{coga} \citep{Rpackage:coga}
for {\bf co}nvolutions of {\bf ga}mma distributions.
However, this computation requires the knowledge of
the eigenvalues $\lambda_1,\ldots,\lambda_p$
that we are trying to estimate.
Therefore we need a transformation of the Euclidean
distances such that the transformed distances have
an approximate distribution whose quantiles do not
require knowing $\lambda_1,\ldots,\lambda_p$\;.

In the simplest case $\lambda_1=\ldots=\lambda_p$
(constant eigenvalues), and then $||X||^2/\lambda_1$ 
follows a $\chi_p^2$ distribution.
It is known that when $p$ increases the distribution
of $||X||^2$ tends to a Gaussian distribution, but 
this also holds for some other powers of $||X||$.
\cite{WilsHilf} found that the best transformation of
this type was $||X||^{2/3}$ in the sense of coming
closest to a Gaussian distribution.
The quantiles $q_\alpha$ of a Gaussian distribution
are easier to compute and can then be transformed 
back to $q_\alpha^{3/2}$.

\begin{figure}[!ht]
\centering
\vspace{0.2cm}
\includegraphics[width=0.89\textwidth]
	{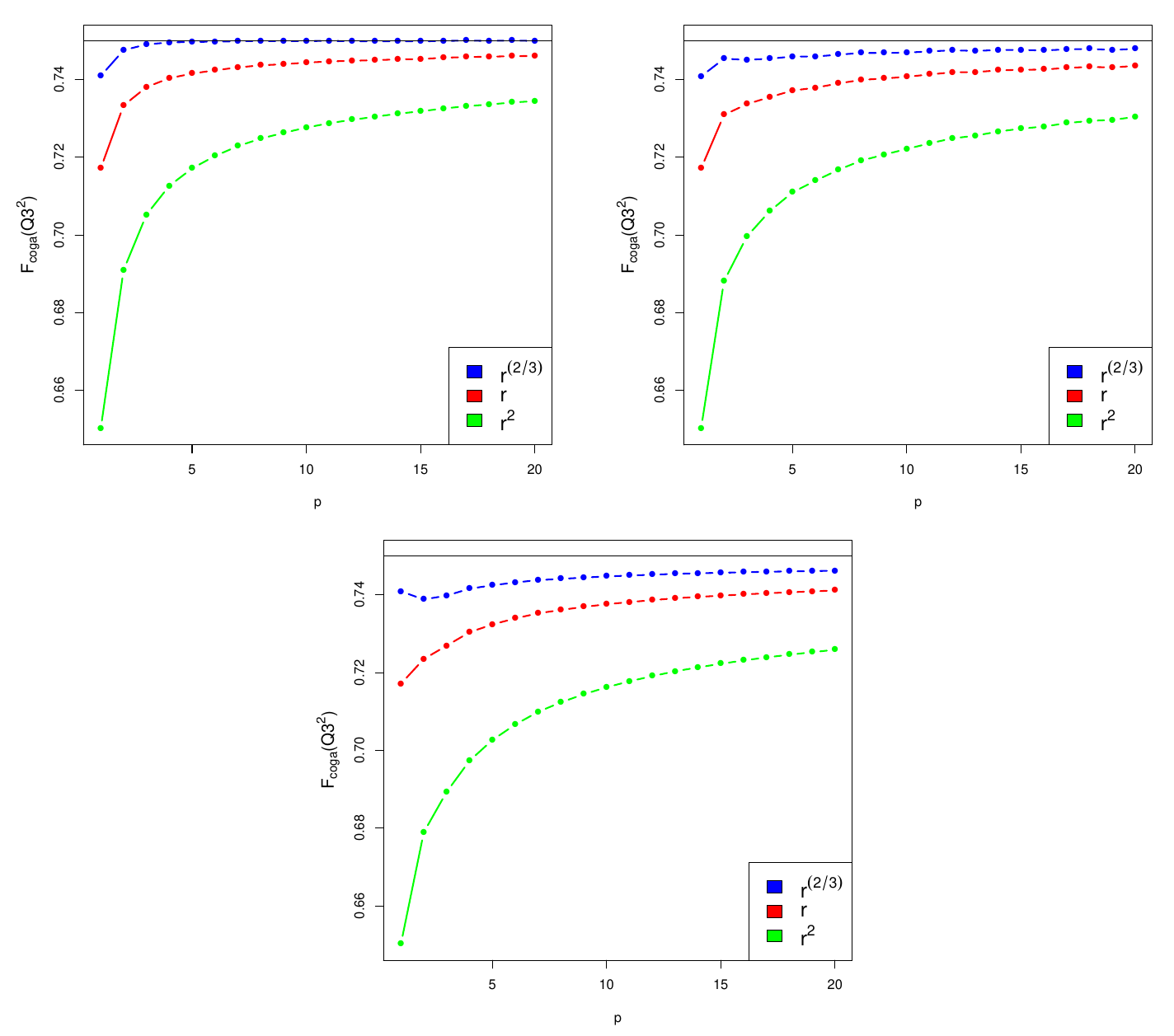} 
\caption{Approximation of the third quartile of a
  coga distribution for dimensions $p=1,\ldots,20$
	when the eigenvalues are constant (top left),
	linear (top right), or quadratic (bottom), using
	three different normalizing transforms.}
\label{fig:approx}
\end{figure}

It turns out that the same Wilson-Hilferty
transformation also works quite well in the more 
general situation where the eigenvalues  
$\lambda_1,\ldots,\lambda_p$ need not be the same.
We came to this conclusion by a simulation study,
a part of which is illustrated here.
The dimension $p$ ranged from 1 to 20 by steps of 1.
For each $p$ we generated $n = 10^6$ observations 
$y_1,\ldots,y_n$ from the coga distribution with 
shape parameters $(0.5,\ldots,0.5)$.
The scale parameters had three settings:
constant $(2,2,\ldots,2)$, linear 
$(p,(p-1),\ldots,1)$, and quadratic
$(p^2, (p-1)^2, \ldots, 1)$, after which the 
scale parameters were further standardized in 
order to sum to $2p$. 
These correspond to the distribution of the squared 
Euclidean norms of a multivariate normal distribution 
where the covariance matrix has eigenvalues that
are constant or proportional to $(p,(p-1),\ldots,1)$ 
(linear eigenvalues) or to $(p^2,(p-1)^2,\ldots,1)$ 
(quadratic eigenvalues). 
Denote the unsquared Euclidean norms as 
$r_i = \sqrt{y_i}$.
Then we estimate quantiles, e.g. $Q_3$ by assuming 
normality of the transformed values $h_1(r_i)=r_i^2$
(square),  $h_2(r_i)=r_i$ (Fisher), and 
$h_3(r_i)=r_i^{2/3}$ (Wilson-Hilferty), by computing
the third quartile of a gaussian distribution with
$\hmu = \med_i(h(r_i))$ and $\hs = \mad_i(h(r_i))$.
Finally, we have evaluated the cumulative 
distribution function of the coga distribution in 
$\hat{Q}_3^2$. Ideally, we would like to 
obtain $F_{\mbox{coga}}(\hat{Q}_3^2) = 0.75$.
The result of this experiment is shown in Figure
\ref{fig:approx}.
We clearly see that the Wilson-Hilferty transform
brings the approximate quantile closest to its
target value. The results for the first quartile Q1
(not shown) are very similar.

\subsection{Proof of Proposition \ref{prop:eigvec}}
\label{A:eigenvectors}
\noindent
\textbf{Part 1: Preservation of the eigenvectors.}

First note that $g$ is orthogonally equivariant, 
i.e. $g(HX) = Hg(X)$ for any orthogonal matrix $H$.
Therefore $S_g = E_{F_X}[g(X)g(X)^T]$ implies
$E_{F_X}[g(HX)g(HX)^T] = H S_g H^T$.

The distribution of $Z$ is spherically symmetric
hence invariant to reflections along a coordinate 
axis, which are described by diagonal matrices 
$R$ with an entry of -1 and all other entries +1.
For every reflection matrix $R$ it thus holds that
$E_{F_Z}[g(DZ)g(DZ)^T] = E_{F_Z}[g(DRZ)g(DRZ)^T] = 
 E_{F_Z}[g(RDZ)g(RDZ)^T] = RE[g(DZ)g(DZ)^T]R^T$, 
where the third equality holds because $DR=RD$ 
as both $D$ and $R$ are diagonal, and the last
equality because R is orthogonal.
Therefore $E_{F_Z}[g(DZ)g(DZ)^T]$ is a diagonal
matrix, which we can denote as 
$\Lambda_g := \mbox{diag}(\lambda_{g,1},\ldots, 
\lambda_{g,p})$.

Now take $U$ an arbitrary orthogonal 
matrix and let $X = UDZ$. Then\\ 
$S_g = E_{F_Z}[g(UDZ)g(UDZ)^T] = 
 UE_{F_Z}[g(DZ)g(DZ)^T]U^T = U\Lambda_g U^T$.
For the plain covariance matrix $\Sigma$ of X we 
have $\Sigma = E_{F_Z}[UDZ(UDZ)^T] = U\Lambda U^T$
where $\Lambda = D D^T = \mbox{diag}
(\delta_1^2,\ldots,\delta_p^2)$.
Therefore, the same matrix $U$ orthogonalizes 
both $\Sigma$ and $S_g$\,, hence $S_g$ and $\Sigma$ 
have the same eigenvectors.\\

\noindent
\textbf{Part 2: Preservation of the ranks of the 
eigenvalues.}

Let $i>j$ and suppose that $\delta_i>\delta_j$. 
We will show that $\lambda_{g,i}> \lambda_{g,j}$.
Note that 
\begin{equation*}
  \lambda_{g,i} = \int{g(DZ)_i^2f_Z(Z) dZ} = 
  \int{\delta_i^2z_i^2\xi(||DZ||)^2f_Z(Z) dZ},
\end{equation*} 
where $f_Z$ is the density of $Z$.
Similarly, we have 
\begin{equation*}
  \lambda_{g,j} = \int{g(DZ)_j^2f_Z(Z) dZ} =
  \int{\delta_j^2z_j^2\xi(||DZ||)^2f_Z(Z) dZ}.
\end{equation*}
This means that $\lambda_{g,i}> \lambda_{g,j}$ 
is equivalent to:
\begin{equation}\label{eq:prev}
  \int{(\delta_i^2z_i^2-\delta_j^2z_j^2)
	\xi(||DZ||)^2f_Z(Z) dZ > 0}.
\end{equation}
As $Z$ is spherically symmetric, 
i.e. $f_Z(Z)\sim w(||Z||)$, we can write 
\eqref{eq:prev} as 
\begin{equation}\label{eq:A4}
  \int{(\delta_i^2z_i^2-\delta_j^2z_j^2)
	\xi(||DZ||)^2w(||Z||) dZ > 0}\;.
\end{equation}

Note that we can change the variable of integration 
as follows. Let $y_k = \delta_k z_k$ and write 
$Y = (y_1,\ldots,y_p)$.
Then \eqref{eq:A4} is equivalent to 
\begin{equation}\label{eq:A5}
  \frac{1}{\delta_1\cdots \delta_p}\left\{
	\int{(y_i^2-y_j^2) \xi(||Y||)^2	w\left(
	\sqrt{\sum_{k=1}^{p}{\frac{y_k^2}{\delta_k^2}}}
	\right) dY}\right\}>0\;.
\end{equation}
We can ignore the positive constant
$1/(\delta_1\cdots\delta_p)$ and split the
integral over the domains 
$A := \{x\in \mathbb{R}^d| \; |x_i|>|x_j|\}$ 
and $B := \{x\in \mathbb{R}^d| \; |x_i|<|x_j|\}$,
yielding 
\begin{align*}
  & \int{(y_i^2-y_j^2)\xi(||Y||)^2 
	  w\left(\sqrt{\sum_{k=1}^{p}{\frac{y_k^2}
	  {\delta_k^2}}}\right) dY}\\
 =& \int_{A}{(y_i^2-y_j^2)\xi(||Y||)^2 
    w\left(\sqrt{\sum_{k=1}^{p}{\frac{y_k^2}
		{\delta_k^2}}}\right) dY}+ 
		\int_{B}{(y_i^2-y_j^2)\xi(||Y||)^2 
		w\left(\sqrt{\sum_{k=1}^{p}{\frac{y_k^2}
		{\delta_k^2}}}\right) dY}\\
 =& \int_{A}{(y_i^2-y_j^2)\xi(||Y||)^2
    w\left(\sqrt{\sum_{k=1}^{p}{\frac{y_k^2}
		{\delta_k^2}}}\right) dY}+ 
		\int_{A}{(y_j^2-y_i^2)\xi(||Y||)^2 
		w\left(\sqrt{\sum_{k=1}^{p}{\frac{y_k^2}
		{\delta_k^2}}+\Delta_{ij}}\right) dY}\\
 =& \int_{A}{(y_i^2-y_j^2)\xi(||Y||)^2
    \left\{w\left(\sqrt{\sum_{k=1}^{p}
		{\frac{y_k^2}{\delta_k^2}}}\right)-
		w\left(\sqrt{\sum_{k=1}^{p}{\frac{y_k^2}
		{\delta_k^2}}+\Delta_{ij}}\right)\right\}dY}
\end{align*}
where in the second equality we have changed the
variables of the integration over $B$ by replacing
$(y_i,y_j)$ by $(-y_j,y_i)$ which has Jacobian 1.
The $\Delta_{ij}$ in that step is the correction
term $\Delta_{ij} = 
\frac{y_i^2}{\delta_j^2} + \frac{y_j^2}{\delta_i^2}
-\frac{y_i^2}{\delta_i^2} -\frac{y_j^2}{\delta_j^2} 
= \frac{y_i^2-y_j^2}{\delta_j^2}
 -\frac{y_i^2-y_j^2}{\delta_i^2} 
= (y_i^2-y_j^2)\left(\frac{1}{\delta_j^2}-
  \frac{1}{\delta_i^2}\right)$.

Note that on $A$ it holds that $|y_i|>|y_j|$ hence
$y_i^2-y_j^2>0$ so $\Delta_{ij}>0$. Since
$w$ is a decreasing function it follows that
\begin{equation}\label{eq:condit}
  w\left(\sqrt{\sum_{k=1}^{p}{\frac{y_k^2}
  {\delta_k^2}}}\right)-w\left(\sqrt{\sum_{k=1}^{p}
	{\frac{y_k^2}{\delta_k^2}}+\Delta_{ij}}\right)>0
\end{equation}
which implies \eqref{eq:A5} so 
$\lambda_{g,i}> \lambda_{g,j}$\,.
If on the other hand $\delta_i$ and $\delta_j$
are tied, i.e. $\delta_i=\delta_j$\,, it follows
that  $\Delta_{ij}=0$ hence
$\lambda_{g,i}=\lambda_{g,j}$\,.

\subsection{Influence function}
\label{A:IF}
\noindent
\textbf{Proof of Proposition \ref{prop:IF}.}

Consider the contaminated distribution 
$F_{\eps, z} = (1-\eps)F + \eps \Delta_z$
where $z\in \mathbb{R}^p$ and $\eps \in [0,1]$. 
We then have:
\begin{align*}
  S(F_{\eps, z}) & = E_{F_{\eps, z}} 
  \left[ g(X)g(X)^T \right] \\
& = (1 - \eps) \int{g_{\eps}(X)g_{\eps}(X)^T dF(X)}
  + \eps \int{g_{\eps}(X)g_{\eps}(X)^T d\Delta_z}\;.
\end{align*}
If we take the derivative with respect to $\eps$ 
and evaluate it in $\eps = 0$, we get:
\begin{align*}
  \left. \frac{\partial}{\partial \eps} S(F_{\eps, z}) 
	\right|_{\eps = 0} = g(z)g(z)^T - \Xi_g + 
	\left. \frac{\partial}{\partial \eps} 
	\int{g_{\eps}(X)g_{\eps}(X)^T dF(X)} 
	\right|_{\eps = 0}\;.\\
\end{align*}

\noindent
\textbf{Calculation of the IF.}

While the expression of the influence function might 
seem relatively simple, its (numerical) calculation 
is rather involved. We can write:
\begin{align*}
  & \left. \frac{\partial}{\partial \eps} 
	  \int{g_{\eps}(X)g_{\eps}(X)^T dF(X)} 
	  \right|_{\eps = 0}\\
 =& \left.\int{ \frac{\partial}{\partial \eps} 
    \left(g_{\eps}(X) \right) g_{\eps}(X)^T + 
		g_{\eps}(X) \frac{\partial}{\partial \eps}  
		\left(g_{\eps}(X)^T\right) dF(X)}
		\right|_{\eps = 0}\\
 =& \int{ \left(\left.\frac{\partial}{\partial \eps} 
    g_{\eps}(X) \right|_{\eps = 0}\right) g(X)^T + 
		g(X) \left(\left.\frac{\partial}{\partial \eps} 
		g_{\eps}(X)^T\right|_{\eps = 0}\right) dF(X)}.
\end{align*}
So the term we need to determine is 
$\left.\frac{\partial}{\partial \eps} 
g_{\eps}(X) \right|_{\eps = 0}$.
Recalling that $g(t) = t \xi(||t||)$ we have 
$g_{\eps}(t) = t \xi_{\eps}(||t||)$. 
This means that the contamination affects $g$ 
because it affects the radial function $\xi$. 
Therefore we have to compute
$\left.\frac{\partial}{\partial \eps} 
g_{\eps}(X) \right|_{\eps = 0} = 
X\left.\frac{\partial}{\partial \eps} 
\xi_{\eps}(||X||) \right|_{\eps = 0}$
for the functions $g$ given by 
\eqref{eq:Winsor}--\eqref{eq:LR}.

In these functions $\xi$ depends on $F_X$ though the 
distribution of $||X||^{2/3}$.
Suppose that $||X||^{2/3}\sim G$ when $X\sim F$, so
$G$ is a univariate distribution.
For $X_{\eps} \sim F_{\eps, z} =
 (1-\eps)F + \eps \Delta_z$ we then have 
$||X_{\eps}||^{2/3} \sim G_{\eps, ||z||^{2/3}}
 = (1-\eps)G + \eps \Delta_{||z||^{2/3}}$\;.
For uncontaminated data the density of $||X||^{2/3}$ 
is given by 
\[f_{G}(t) = f_{\mbox{coga}}(t^3)|3t^2|\;,\]
where $f_{\mbox{coga}}$ is the density of the 
{\bf co}nvolution of {\bf ga}mma distributions. 
We need this density to evaluate the influence 
functions of their median and mad.

The cutoffs in the paper are
\begin{align*}
Q_1 &=& \left(\mbox{hmed}{||X||^{\frac{2}{3}}} - 
        \mbox{hmad}{||X||^{\frac{2}{3}}}
				\right)^{\frac{3}{2}}	\\
Q_2 &=& \left(\mbox{hmed}{||X||^{\frac{2}{3}}}
        \right)^{\frac{3}{2}}\\
Q_3 &=& \left(\mbox{hmed}{||X||^{\frac{2}{3}}} + 
        \mbox{hmad}{||X||^{\frac{2}{3}}}
				\right)^{\frac{3}{2}}\\
Q_3^* &=& \left(\mbox{hmed}{||X||^{\frac{2}{3}}} + 
        1.4826\;\mbox{hmad}{||X||^{\frac{2}{3}}}
				\right)^{\frac{3}{2}}
\end{align*}
and we can compute their influence functions:
\begin{align*}
\IF(z, Q_1,F) =&\; \frac{3}{2} \sqrt{\med(G) - 
  \; \mad(G)} \; (\IF(||z||^{2/3}, \med, G)- \; 
	\IF(||z||^{2/3}, \mad, G))\\
\IF(z, Q_2,F) =&\; \frac{3}{2} \sqrt{\med(G)}\;
  \IF(||z||^{2/3}, \med, G)\\
\IF(z, Q_3,F) =&\; \frac{3}{2} \sqrt{\med(G) + \; 
  \mad(G)} \; (\IF(||z||^{2/3}, \med, G)+ \; 
	\IF(||z||^{2/3}, \mad, G))\\
\IF(z, Q_3^*,F) =&\; \frac{3}{2} \sqrt{\med(G) +
  1.4826 \; \mad(G)} \; (\IF(||z||^{2/3}, \med, G)\\ 
  &\; +1.4826 \; \IF(||z||^{2/3}, \mad, G)).
\end{align*}
The Winsor GSSCM is given by
$\xi(r) = \mathbbm{1}_{r \ls Q_2}+
 \frac{Q_2}{r}\mathbbm{1}_{r > Q_2}$.
For the contaminated case this becomes 
$\xi_{\eps}(r) = 
 \mathbbm{1}_{r \ls Q_{2,\eps}} + 
 \frac{Q_{2,\eps}}{r}\mathbbm{1}_{r > Q_{2,\eps}}$.
We then have:
\begin{align*}
\frac{\partial}{\partial \eps}\xi_{\eps}(r) &= 
  \frac{\partial}{\partial \eps}\left\{
	\mathbbm{1}_{[0, Q_{2,\eps}]}(r) + 
	\frac{Q_{2,\eps}}{r}\mathbbm{1}_{(Q_{2,\eps}, 
	\infty)}(r)\right\}\\
 &= \delta(r-Q_{2,\eps}) Q_{2,\eps}' + 
  \frac{Q_{2,\eps}'}{r}  \mathbbm{1}_{(Q_{2,\eps}, 
	\infty)}(r)- \frac{Q_{2,\eps}}{r} 
	\delta (r-Q_{2,\eps}) Q_{2,\eps}'\;,
\end{align*}
where $\delta(x - y)$ denotes the distributional 
derivative of $\mathbbm{1}_{(-\infty,x]}(y)$.
Evaluation in $\eps = 0$ gives
\begin{align*}
 \delta(r-Q_2) IF(z,Q_2,F) +
 \frac{\IF(z,Q_2,F)}{r}
 \mathbbm{1}_{(Q_2, \infty)}(r)- \frac{Q_2}{r} 
 \delta (r-Q_2)IF(z,Q_2,F)\\
= \left(1 - \frac{Q_2}{r}\right)\delta(r-Q_2)
 IF(z,Q_2,F)+ \frac{\IF(z,Q_2,F)}{r} 
 \mathbbm{1}_{(Q_2,\infty)}(r)\;.
\end{align*}
As $\left(1 - \frac{Q_2}{r}\right)\delta(r-Q_2)$ is 0 
everywhere, we only need to integrate the last term. 
This yields
\begin{align*}
 \left.\frac{\partial}{\partial \eps} g_{\eps}(X) 
 \right|_{\eps = 0} = \frac{X}{||X||}\IF(z,Q_2,F) 
 \mathbbm{1}_{(Q_2, \infty)}(||X||)\;.
\end{align*}
The influence function of $S_g$ is thus given by:
\begin{align*}
\IF(z,S_g,F) =& g(z)g(z)^T - \Xi_g(F)\\
 &+ \int{ \left(\frac{X }{||X||}\IF(z,Q_2,F) 
  \mathbbm{1}_{(Q_2, \infty)}(||X||)\right)
	g(X)^T dF(X)}\\
 &+ \int{g(X) \left(\frac{X }{||X||}\IF(z,Q_2,F)
  \mathbbm{1}_{(Q_2, \infty)}(||X||)\right)^T dF(X)}.
\end{align*}
Note that the last 2 terms in the sum are each other's 
transpose. The integration is done numerically.\\

The derivation of the influence function of the 
Quad GSSCM is entirely similar to that of Winsor. 
The main difference is that now 
$\left.\frac{\partial}{\partial \eps} 
g_{\eps}(X) \right|_{\eps = 0}$ is given by
\begin{align*}
  \left.\frac{\partial}{\partial \eps} g_{\eps}(X) 
  \right|_{\eps = 0}
  = 2 Q_2 \IF(z,Q_2,F)\frac{X }{||X||^2} 
	\mathbbm{1}_{(Q_2, \infty)}(||X||)\;.\\
\end{align*}

The linearly redescending (LR) method uses a second 
cutoff:
\begin{equation}
\displaystyle
  \xi(r) = 
	\begin{cases}
	1 & \mbox{ if } r \ls Q_2 \\
	(Q_3^* - r)/(Q_3^* - Q_2) & 
	  \mbox{ if } Q_2 < r \ls Q_3^* \\
  0 & \mbox{ if } r > Q_3^*\;.
   \end{cases}
\end{equation}
In the contaminated case we obtain 
$g_{\eps}(x) = x \xi_{\eps}(||x||)$ with
\begin{equation}
\displaystyle
  \xi_{\eps}(r) = 
	\begin{cases}
	 1 & \mbox{ if } r \ls Q_{2, \eps} \\
	 (Q_{3,\eps}^* - r)/(Q_{3,\eps}^* - Q_{2,\eps})
	  & \mbox{ if }
		Q_{2,\eps} < r \ls Q_{3,\eps}^* \\
   0 & \mbox{ if } r > Q_{3,\eps}^*\;.
   \end{cases}
\end{equation}
Taking the derivative with respect to $\eps$ yields:
\begin{align*}
  \frac{\partial}{\partial \eps}\xi_{\eps}(r) 
	&= \delta (r-Q_{2,\eps}) + 
	  \frac{Q_{3, \eps}^* - r}{Q_{3,\eps}^* - 
		Q_{2,\eps}}\left(\delta (r-Q_{3,\eps}^*)- \delta (r-Q_{2,\eps})\right)\\
  &+ \mathbbm{1}_{[Q_{2,\eps},Q_{3,\eps}^*]}
	  \frac{Q_{3,\eps}^{*\prime}(Q_{3,\eps}^*-Q_{2,\eps})
		- (Q_{3,\eps}^{*\prime} - Q_{2,\eps}')
		(Q_{3,\eps}^*-r)}{(Q_{3,\eps}^* - Q_{2,\eps})^2}\;.
\end{align*}
Evaluation in $\eps = 0$ gives:
\begin{align*}
 & \delta (r-Q_{2}) +
    \frac{Q_{3}^* - r}{Q_{3}^* - Q_{2}}\left(
		\delta (r-Q_{3}^*) - \delta (r-Q_{2})\right)\\
 &+ \mathbbm{1}_{[Q_{2},Q_{3}^*]}
    \frac{\IF(z,Q_3^*,F)(Q_{3}^*-Q_{2}) - 
		(\IF(z,Q_3^*,F) - \IF(z,Q_2,F))(Q_{3}^*-r) }
		{(Q_{3}^* - Q_{2})^2}\;.
\end{align*}	
When integrating only the last term plays a role,
yielding
\begin{align*}
  \left.\frac{\partial}{\partial \eps} g_{\eps}(X) 
	\right|_{\eps = 0}
 =& X \mathbbm{1}_{[Q_{2},Q_{3}^*]}(||X||)\\
  & \frac{\IF(||z||,Q_3^*,F)(Q_{3}^*-Q_{2}) - 
	  (\IF(||z||,Q_3^*,F) -
		\IF(||z||,Q_2,F))(Q_{3}^*-||X||) }
		{(Q_{3}^* - Q_{2})^2}\\
=& X \mathbbm{1}_{[Q_{2},Q_{3}^*]}(||X||)
  \frac{\IF(||z||,Q_3^*,F)(||X||-Q_{2}) + 
   \IF(||z||,Q_2,F)(Q_{3}^*-||X||)}
	 {(Q_{3}^* - Q_{2})^2}\;.
\end{align*}

For the Ball SSCM we analogously derive that
\begin{align*}
  \left.\frac{\partial}{\partial \eps} g_{\eps}(X) 
	\right|_{\eps = 0}
  = \delta(||X||-Q_2)\IF(z, Q_2, F)X\;.
\end{align*}

\vskip0.2cm
Finally, for the Shell SSCM we obtain
\begin{align*}
 \left.\frac{\partial}{\partial \eps} g_{\eps}(X)
 \right|_{\eps = 0}
 = \left(\delta(||X||-Q_3) \IF(z, Q_3, F)
  -\delta(||X||-Q_1)
	\IF(z, Q_1, F)\right)X\;.\\
\end{align*}

\newpage
\subsection{Breakdown values}
\label{A:breakdown}
\noindent
\textbf{Proof of Proposition \ref{prop:bdvS}.}
\begin{proof}
Denote by $\mathcal{J}$ the set of all subsets of 
$\{1, \ldots, n\}$ with $p+1$ elements.
For every subset $J \in \mathcal{J}$ we define 
$\displaystyle \eta_J \coloneqq \max_{i\in J}
{d^2(x_i, H_J)}$, where $H_J$ is the hyperplane 
minimizing 
$\displaystyle \sum_{i\in J}{d^2(x_i, H)}$ over all 
possible hyperplanes $H$ and $d(x,H)$ is the 
Euclidean distance between a point $x$ and a 
hyperplane $H$. 
Define $\displaystyle \eta_{\bX} \coloneqq 
\min_{J \in \mathcal{J}}{\eta_J}$. 
Since the original points $\{x_1, \ldots, x_n\}$ 
are in general position, no $p+1$ points can lie on 
the same hyperplane, which ensures that
$\eta_{\bX} > 0$.
We also put 
$c_1 \coloneqq \max_i||x_i - T(\bX)|| < \infty$.\\ 

\textbf{Part 1.} We first need to show that 
$\epsilon^* \gs \left\lfloor (n-p+1)/2
 \right\rfloor /n$.

Let $m < \left\lfloor (n-p+1)/2\right\rfloor$ 
and replace  
$m$ observations of $\bX = \{x_1,\ldots, x_n\}$ 
yielding $\bX^*$ with location estimate $T(\bX^*)$.
Because $\frac{m}{n}$ is below the breakdown value 
of $T$, there is a constant $c_2 < \infty$ so that 
$||T(\bX^*)-T(\bX)|| \ls c_2$ for all such 
contaminated datasets $\bX^*$. 
By the triangle inequality 
$||x_i - T(\bX^*)|| \ls c_1 + c_2 < \infty$. 
This implies $\hmed(d_i^*) \ls c_1 + c_2$\,, 
hence $\hmed(d_i^*) + 1.4826\,\hmad(d_i^*) 
\ls 2.4826 \; \hmed(d_i^*) \ls 
2.4826(c_1 + c_2)$, where $d_i^*=||x_i^*-T(\bX^*)||$. 
Therefore $||g(t)|| \ls 2.4826(c_1+c_2)$ by 
condition 3. 

First we show that the largest eigenvalue of 
$S_g(\bX^*)$ is bounded over all such datasets 
$\bX^*$. 
Take any $\bX^*$, obtained by replacing $m$ points 
of $\bX$ by arbitrary points. Then
\begin{align*}
\lambda_{\mbox{max}} 
 &= \sup_{||u|| = 1}{u^T S_g(\bX^*) u} = 
  \sup_{||u|| = 1}\frac{1}{n}\sum_{i=1}^{n}
	{u^T g(x_i^*-T(\bX^*))g(x_i^*-T(\bX^*))^T u}\\
 &= \sup_{||u|| = 1}\frac{1}{n}\sum_{i=1}^{n}
  {\left(u^Tg(x_i^*-T(\bX^*))\right)^2} \ls 
	\sup_{||u|| = 1}\frac{1}{n}\sum_{i=1}^{n}
	{||u||^2 ||g(x_i^*-T(\bX^*))||^2}\\
&\ls (2.4826(c_1 + c_2))^2 < \infty.
\end{align*}

Next we show that the smallest eigenvalue of 
$S_g(\bX^*)$ has a positive lower bound 
for all contaminated datasets $\bX^*$.
By condition 2 on $\xi$ we know that 
$\#\{x_i | \;\xi(||x_i-T(\bX^*)||) = 1\} \gs 
\left\lfloor (n + p + 1)/2 \right\rfloor$. 
Therefore, we have at least 
$\left\lfloor (n+p+1)/2 \right\rfloor - 
(\left\lfloor (n-p+1)/2\right\rfloor -1) = p+1$ 
regular points for which 
$\xi(||x_i-T(\bX^*)||) = 1$, let's assume w.l.o.g. 
that these are $x_1,\ldots,x_{p+1}$.
We can now write
\begin{align*}
\lambda_{\mbox{min}}
  = \min_{||u|| = 1}{u^T S_g(\bX^*) u}
 &= \min_{||u|| = 1}\frac{1}{n}\sum_{i=1}^{n}
    {u^T g(x_i^*-T(\bX^*))g(x_i^*-T(\bX^*))^T u}\\
 &= \min_{||u|| = 1}\frac{1}{n}\sum_{i=1}^{n}
   {\left(u^Tg(x_i^*-T(\bX^*))\right)^2}\\
 &\gs \min_{||u|| = 1}\frac{1}{n}\sum_{i=1}^{p+1}
   {\left(u^T (x_i-T(\bX^*)) 
	  \xi(x_i-T(\bX^*))\right)^2}\\
&= \min_{||u|| = 1}\frac{1}{n}\sum_{i=1}^{p+1}
   {\left(u^T (x_i-T(\bX^*))\right)^2}\\
&\gs \frac{1}{n}\sum_{i=1}^{p+1}{d^2(x_i, 
  H_{\{1,\ldots,p+1\}})} \gs \eta_{\bX} >0\;.\\
\end{align*}

\textbf{Part 2.} It remains to show that  
$\epsilon^* \ls \left\lfloor (n-p+1)/2
\right\rfloor /n$. 
This is the known upper bound for affine equivariant 
scatter estimators but that result doesn't apply here, 
so we need to show it for this case.
Take any $m \gs \left\lfloor (n-p+1)/2 \right\rfloor$ 
and replace the last $m$ points of $\bX$, keeping the
points $x_1,\ldots,x_{n-m}$ unchanged.
By location equivariance we can assume w.l.o.g. that 
the average of $x_1,\ldots,x_{n-m}$ is zero.
For $j = n-m+1,\ldots,n$ put $x_j^* = \lambda a_j$ 
where $a_j$ is such that
$\min_{i=n-m+1,\ldots,n}{||a_j - a_i||} \gs 1$ and 
such that for all  $\lambda > 1$ it holds that
$\min_{i=1,\ldots,n-m}{||\lambda a_j - x_i||} 
\gs \lambda$. This is possible by placing the $a_j$ 
outside of the convex hull of $\bX$ and far enough 
from each other and $\bX$.

Now consider an unbounded increasing sequence of 
$\lambda_k > 1$. For every $\lambda_k$ the set 
$\{x_{n-m+1}^*,\ldots,x_n^*\}$ must contain at least
one point for which $w_i = 1$, call this point $x_b^*$. 
Take another point of $\bX^*$ for which $w_i = 1$, 
name this $x_c^*$. Note that $x^*_c$ can be an 
original data point or a replaced point.
We now have that $||x_b^* - x_c^*|| \gs \lambda$ hence
$||x_b^*-T(\bX^*)|| + ||x_c^*-T(\bX^*)|| \gs \lambda$.
Therefore $||x_b^*-T(\bX^*)||^2 + ||x_c^*-T(\bX^*)||^2 
\gs \lambda^2/2$.
We then obtain
\begin{align*}
  \sum_{j=1,\ldots,p} \lambda_j(S(\bX^*)) 
  &= \mbox{trace}(S(\bX^*)) \\ 
	&= \frac{1}{n} \sum_{i=1,\ldots,n} \mbox{trace}
	    ((x_i^*-T(\bX^*))(x_i^*-T(\bX^*))^T)\\
  &= \frac{1}{n} \sum_{i=1,\ldots,n}
	  ||x_i^* - T(\bX^*)||^2\\
  &\gs \frac{1}{n} (||x_b^*-T(\bX^*)||^2 +
	  ||x_c^*-T(\bX^*)||^2)\\
  &\gs \lambda^2/(2n)\;.
\end{align*}
This becomes arbitrarily large and so $S(\bX^*)$ 
explodes.
\end{proof}

\noindent
\textbf{Proof of Proposition \ref{prop:bdvT}.}
\begin{proof}
Showing that 
$\eps^*(T,\bX) \ls \left\lfloor (n+1)/2 \right\rfloor/n$
is easy, since $\left\lfloor (n+1)/2 \right\rfloor/n$ 
is the upper bound on the breakdown value of all 
translation equivariant location estimators, see e.g.
\cite{Lopuhaa:BDP}.

It remains to show that $\eps^*(T, \bX) \gs 
\left\lfloor (n+1)/2 \right\rfloor/n$.

Note that the objective given by the sum of the $h$ 
smallest squared Euclidean distances is nonincreasing 
in every C-step.
The value of the objective function after step $k$ is
$\sum_{j=1}^{h}{d_{(j)}^2(\bX,T_k(\bX))}$ where 
$d_{(j)}(\bX,T_k(\bX))$ denotes the $j$-th order 
statistic of the distances $||x_i-T_k(\bX)||$, and 
we have that 
$\sum_{j=1}^{h}{d_{(j)}^2(\bX,T_k(\bX))} \ls 
\sum_{j=1}^{h}{d_{(j)}^2(\bX,T_{k-1}(\bX))}$.

Recall that $h = \left\lfloor (n+1)/2 \right\rfloor$
and define $c_1 \coloneqq \max_{i}{||x_i - T_k(\bX)||} 
< \infty$.
Let $m < n-h$ and replace w.l.o.g. the last $m$ 
observations of $\bX = \{x_1,\ldots, x_n\}$ to obtain 
$\bX^* = \{x_1,\ldots,x_{n-m},x_{n-m+1}^*,\ldots,
x_{n}^*\} = \{x_1^*,\ldots, x_n^*\}$.
Since the spatial median $T_0$ does not yet break down 
for this $m$ \citep{Lopuhaa:BDP}, there is a constant 
$c_2$ such that 
$\max_{i}{||x_i - T_0(\bX^*)||} \ls c_2 < \infty$ for 
all such datasets $\bX^*$.

Consider $T_k(\bX^*)$ and the corresponding objective 
function $\sum_{j=1}^{h}{d_{(j)}^2(\bX^*,T_k(\bX^*))}$. 
Since the C-step does not increase the value of the 
objective function, we have that 
\begin{equation*}
 \sum_{j=1}^{h}{d_{(j)}^2(\bX^*,T_k(\bX^*))} \ls 
 \sum_{j=1}^{h}{d_{(j)}^2(\bX^*,T_{k-1}(\bX^*))} 
 \ls \ldots 
 \ls \sum_{j=1}^{h}{d_{(j)}^2(\bX^*,T_0(\bX^*))}.
\end{equation*}
Note that 
\begin{align*}
 \sum_{j=1}^{h}{d_{(j)}^2(\bX^*,T_0(\bX^*))}
 &\ls \sum_{i=1}^{h}{||x_i^* - T_0(\bX^*)||^2} = 
  \sum_{i=1}^{h}{||x_i - T_0(\bX^*)||^2} \\
 &\ls \left(\sum_{i=1}^{h}{||x_i - T_0(\bX^*)||}
  \right)^2 \ls \left( hc_2\right)^2.
\end{align*}

Since $m$ is at most 
$\left\lfloor (n-1)/2 \right\rfloor$ and 
$h = \left\lfloor (n+1)/2 \right\rfloor	$ we have at 
least $\left\lfloor (n+1)/2 \right\rfloor -
 \left\lfloor (n-1)/2 \right\rfloor = 1$ 
point $x_j$ with $1 \ls j \ls n-m$ for which 
$||x_j - T_k(\bX^*)||^2 \ls d_{(h)}^2(\bX^*,T_k(\bX^*))$. 
Note that $||x_j - T_k(\bX^*)||^2 \ls 
 \sum_{j=1}^{h}{d_{(j)}^2(\bX^*,T_k(\bX^*))} \ls 
 \sum_{j=1}^{h}{d_{(j)}^2(\bX^*,T_0(\bX^*))}$.
So for this $x_j$ we can write
\begin{align*}
 ||T_k(\bX^*) - T_0(\bX)|| 
 &\ls ||T_k(\bX^*) - x_j|| + ||x_j - T_0(\bX) || \\
 & \ls hc_2 + c_1 < \infty\;.
\end{align*} 
Note that this upper bound does not depend on $k$ and 
therefore remains valid when the procedure is iterated 
until convergence ($k \rightarrow \infty$).
\end{proof}

\end{document}